# An Efficient Model Maintenance Approach for MLOps

**Forough Majidi · Foutse Khomh · Heng Li · Amin Nikanjam**

**Abstract** In recent years, many industries have utilized machine learning (ML) models in their systems. Ideally, ML models should be trained on and applied to data from the same distributions. However, the data evolves over time in many application areas, leading to concept drift, which in turn causes the performance of the ML models to degrade over time. Therefore, maintaining up-to-date ML models plays a critical role in the MLOps pipeline. Existing ML model maintenance approaches are often computationally resource-intensive, costly, time-consuming, and model-dependent. Thus, we propose an improved MLOps pipeline, a new model maintenance approach and a Similarity-Based Model Reuse (SimReuse) tool to address the challenges of ML model maintenance. We identify seasonal and recurrent data distribution patterns in time series datasets throughout a preliminary study. Recurrent data distribution patterns enable us to reuse previously trained models for similar distributions in the future, thus avoiding frequent unnecessary retrainings. Then, we integrated the model reuse approach into the MLOps pipeline and proposed our improved MLOps pipeline. Furthermore, we develop SimReuse, a tool to implement the new components of our MLOps pipeline to store models and reuse them for inference of data segments with similar data distributions in

Forough Majidi
Polytechnique Montréall, Québec, Canada
E-mail: forough.majidi@polymtl.ca

Foutse Khomh
Polytechnique Montréall, Québec, Canada
E-mail: foutse.khomh@polymtl.ca

Heng Li
Polytechnique Montréall, Québec, Canada
E-mail: heng.li@polymtl.ca

Amin Nikanjam
Polytechnique Montréall, Québec, Canada
E-mail: amin.nikanjam@polymtl.ca

the future. Our evaluation results on five time series datasets demonstrate that our model reuse approach can maintain the models' performance while significantly reducing maintenance time, costs, and the number of retrainings. Our model reuse approach achieves ML model performance comparable to the best baselines, while reducing the computation time and costs to 1/8th. Therefore, industries and practitioners can benefit from our approach and use our tool to maintain their ML models' performance in the deployment phase to reduce their maintenance time and costs.

## 1 Introduction

Many companies and industries have started using machine learning (ML) in their businesses or replacing their traditional non-ML-based software systems with ML-based ones (Brynjolfsson and Mcafee, 2017). These industries train an ML model using a training dataset and then use it to make predictions. However, data is not stationary in many fields and changes over time due to concept drift. When this happens, the trained model may not work well with the new data, leading to poor performance. As a result, maintaining up-to-date ML models in the production environment is essential, as their performance can degrade over time due to concept drift (Bayram et al., 2022).

MLOps pipelines (e.g., (Spotify Engineering, 2023; Hopsworks, 2023)) are employed to develop, deploy, and maintain ML-based software systems over time (Sculley et al., 2015). Such pipelines typically contain components like model "model performance monitoring" and "trigger" to train a new model when the performance drops below a specific threshold(Google Cloud, 2024). However, using ML in software systems has its own challenges (CircleCI, 2024). For instance, the increasing size of datasets and the growing complexity of ML models make training a new model and maintaining ML-based software systems progressively more difficult, resource-intensive and time-consuming (CircleCI, 2024).

The literature offers several approaches, like periodical model training, to maintain ML-based software systems and ML models (Lyu et al., 2023) in the production environment. Although these approaches aid in maintaining high prediction accuracy and user satisfaction, each method has its own set of limitations, highlighting the need for a more effective and comprehensive solution. For example, periodical model training is one of the best maintenance approaches that mostly achieves high performance compared to other approaches (Lyu et al., 2024, 2021a). While increasing the frequency of model retraining can improve the performance (Lyu et al., 2021a), retraining is costly, time-consuming, and requires expensive hardware resources. To the best of our knowledge, none of the prior works looked at the recurrent and seasonal data distributions to reuse the previously trained model in the future to save the cost and time of ML model maintenance. Furthermore, other model maintenance approaches like online learning (Lyu et al., 2024) are model-dependent and cannot be applied to all types of models. Therefore, this research aims to 1) explore the seasonality and recurrent drifts in the time series data by conducting a preliminary study and 2) provide a novel model-agnostic maintenance approach that maintains ML models in production while keeping resource usage, maintenance time, and cost low, and 3) evaluate the performance and cost of the proposed method. We also developed a tool called SimReuse that helps us save models and use them again later to make predictions on data with a similar data distribution to the training data.

Our study is organized by conducting a preliminary study, proposing our maintenance approach and then answering two Research Questions (RQs) to evaluate our approach:

**Preliminary study**: Time series datasets are common in many application areas like the stock market price (Timescale Contributors, 2024). The preliminary study aims to understand whether recurrent and similar seasonal distribution patterns exist in the time series datasets. The goal is to understand the characteristics of data relevant to model maintenance. First, we analyze the distribution of time series datasets over time to check if we find any recurrent and seasonal distribution patterns. Then, we split the datasets into specific segments and use similarity measurement techniques to find similar distributions over time. As a result, we observed that recurrent patterns exist. Afterward, we attempt to improve the MLOps pipeline (Google Cloud, 2024) by adding new components to benefit from these recurrent patterns for reusing our previously trained models. Our improved MLOps pipeline reuses the previously trained models for similar data distributions in the future. This new model maintenance approach reduces computation resources, time, and maintenance costs. Developers can set the parameters of SimReuse and use this tool for their projects. To assess the effectiveness of SimReuse, we investigate the following research questions:

**RQ1: How does SimReuse perform regarding ML performance metrics?** Keeping the performance high is one of the critical factors for practitioners when choosing an ML model maintenance approach. We evaluate the performance impact of our approach by comparing it to our baseline approaches to ensure it works well in the production environment. The results show that our model reuse approach performs comparably to the periodically updated models and informed retrained models while outperforming the stationary and random output models.

**RQ2: How does SimReuse perform regarding maintenance costs?** Training a model is costly for companies (Lyu et al., 2021a). For instance, if a company uses AWS SageMaker with the "ml.p4de.24xlarge" instance, training a model that takes 10 hours costs 370$, which is quite expensive if the training takes longer. Our approach aims to reduce the cost of model maintenance by reusing the existing models while keeping the performance high. Therefore, in RQ2, we evaluate the cost of our model reuse approach and compare it with the cost of baseline approaches. Our results show that SimReuse reduces

maintenance time and cost to 1/8th and 1/1.13th compared to the periodical retraining and informed model update approaches, respectively.

To the best of our knowledge, SimReuse is the first tool that proposed the following contributions to the existing model maintenance techniques:

- Our improved MLOps pipeline and model reuse approach are the first model maintenance solutions that benefit from recurring data distributions and seasonalities in time series datasets to reuse the previously trained models in the ML-based software systems. Furthermore, our approach is model- and application-agnostic, making it applicable to any ML model and time-based dataset.
- We reduce the frequency of model retraining and associated model maintenance costs by utilizing existing models trained on distributions similar to the upcoming distribution, while maintaining performance comparable to the baseline models.
- We demonstrate the effectiveness of our model maintenance approach from the aspects of ML performance metrics and model maintenance costs using five real-world time series datasets. Our approach not only matches the ML performance of the best baseline models but also greatly lowers the costs associated with maintaining the model.
- We forecast the future distribution and proactively choose the ML model before a new data segment arrives.

The remainder of this paper is organized as follows: Section 2 provides background information. Section 3 reviews the related works and positions our work relative to them. Section 4 conducts a preliminary study on the characteristics of the datasets. Section 5 proposes our improved MLOps pipeline and our model reuse approach. Section 6 explains our evaluation process and presents the results for answering our research questions. Section 7 explains the implications of our work for different groups of people. Section 8 discusses the threats to the validity of our study. Finally, section 9 concludes this work and proposes ideas for future works.

## 2 Background

This section explains the necessary background knowledge.

### 2.1 MLOps pipeline

This research introduces an enhanced MLOps pipeline by incorporating new components. Before introducing our new components, we first present an overview of the MLOps pipeline and briefly describe its key elements.

An MLOps pipeline helps manage machine learning projects by automating critical tasks like getting data ready, training models, and deploying them. It helps teams to work more efficiently and faster by using tools/approaches that automatically check the new data, detect drift, retrain models with new data, test them, and make them ready for use. The main goal is to keep the ML system running smoothly and ensure that models can be updated easily when needed. Mixing machine learning with DevOps methods enables MLOps to make machine learning systems more reliable, efficient, and able to grow (Google Cloud, 2024; RunAI Contributors, 2024). Also, Automated Machine Learning (AutoML) tools can be used to automate one or multiple components of the MLOps pipeline (e.g. model training) (Majidi et al., 2022). Figure 1 shows the MLOps pipeline (Google Cloud, 2024).

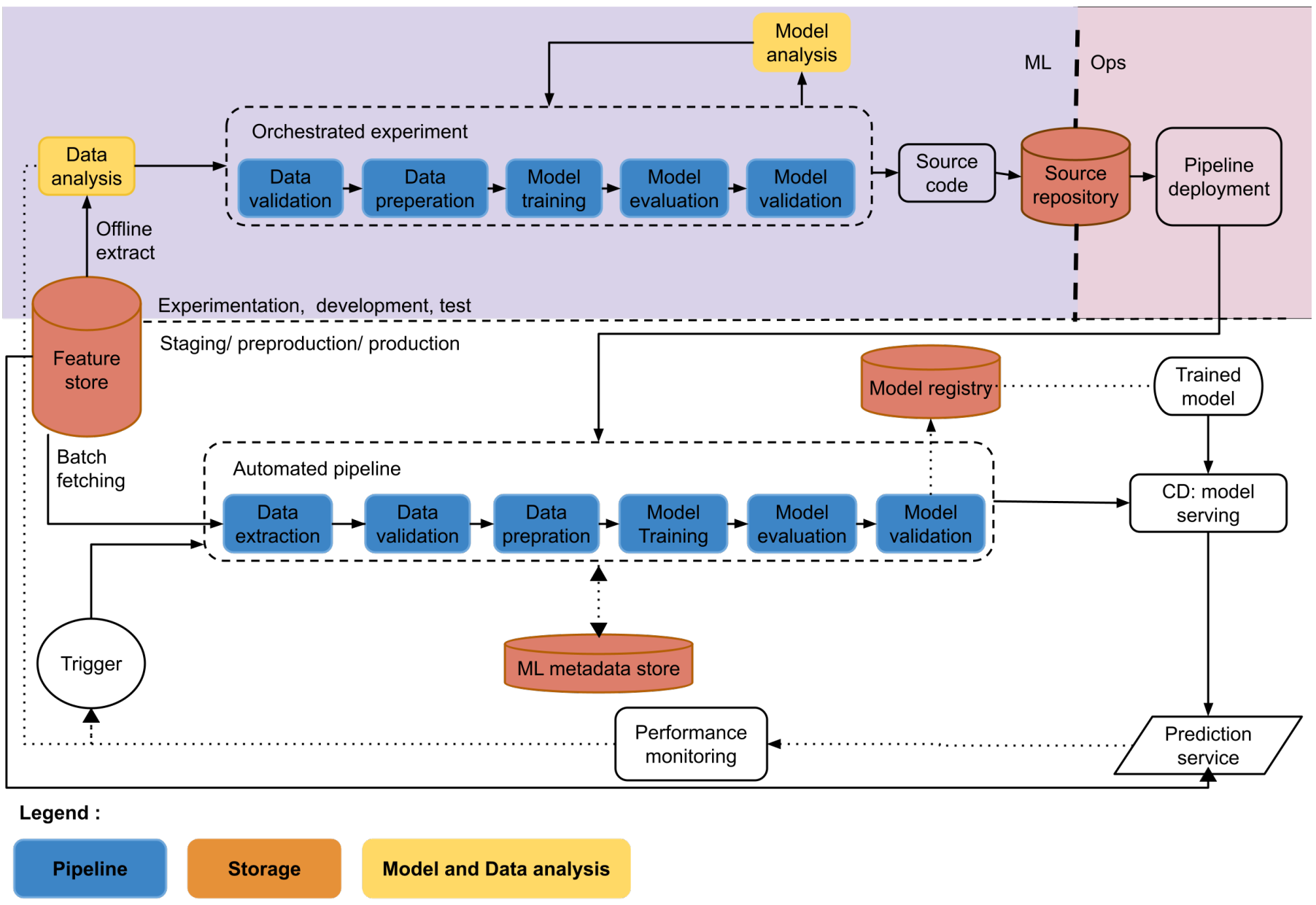

Fig. 1: MLOps pipeline (Google Cloud, 2024)

The explanation of the MLOps components based on the Google definitions (Google Cloud, 2024) is as follows:

**Feature Store**: feature store is an optional component, and it serves as a centralized repository that standardizes the definition, storage, and retrieval of features for both training and serving.

**Data analysis**: Data analysis involves performing exploratory data analysis (EDA) to understand the structure and key details of the data to build the ML model. This step helps identify the necessary data preparation and feature engineering and ensures the data meets the model's expectations.

**Data validation**: This step is needed before training the model to decide if you should retrain the model or stop the pipeline. The pipeline automatically makes this decision if specific issues are found. The issues are Data schema skews and Data values skews.

**Data preparation**: This step is needed before training the model to preprocess data. Preprocessing data includes steps like data transformation, data normalization, data cleaning, and so on. The output of this step is preprocessed data.

**Model training**: In this step, the prepared data is used to apply different algorithms and train various ML models. The algorithm is also fine-tuned using hyperparameter tuning to find the best model. The result of this step is a trained model.

**Model evaluation**: The model is evaluated using the test set data. The output of this step is the results of quality assessment metrics such as the F1 score, accuracy, precision, mean squared error, and so on.

**Model validation**: The model is approved to be suitable for deployment, and its performance is better than baselines.

**Model analysis**: This component provides a deeper inspection of the trained model, extending beyond performance metrics to focus on aspects such as robustness, security, explainability, interpretability, and bias detection.

**Source code**: This component includes the source code.

**Source repository**: The source code is stored in the source repository.

**Pipeline deployment**: The whole training pipeline is deployed. Then, the trained model serves as the prediction service. ML practitioners can deploy their machine learning projects on Docker (Openja et al., 2022).

**Model registry**: The trained model is stored in the model registry.

**Trained model**: This component contains the trained model.

**Model serving**: The validated model is set up in a production environment to provide predictions.

**Data extraction**: The needed data from different sources is combined.

**Prediction service**: The model is used to make predictions on the deployment environment's data.

**ML metadata store**: This component records the important information, including the execution time, parameter arguments, and so on.

**Performance monitoring**: The performance of the ML model in the production environment is checked, and an operation is triggered when the performance drops below a certain threshold.

**Trigger**: When the model's performance in the production environment drops below the threshold, a few operations can be triggered, including 1) going back to the development phase and executing a new experiment cycle and 2) executing the pipeline.

Retraining ML models can be triggered by on-demand execution, on a schedule of routine updates, or occasional updates when new data is available. It may also occur due to declining performance or notable changes in data distributions (concept drift).

## 2.2 Time series data and seasonality

Time series data consists of a collection of observations of a subject over time intervals (Clarify.io Contributors, 2024). Many domains have time series data. For instance, Weather forecasting, demand forecasting, traffic flow prediction, and so on are all time series datasets. Many time series datasets (e.g. economics and business) contain seasonal patterns (Zhang and Qi, 2005). Seasonality refers to recurring changes/distribution patterns that are caused by factors like days of the week, weather, holidays, months of the year, and so on (Zhang and Qi, 2005). This study proposes that if distribution patterns in the dataset repeat, we can reuse models trained on similar data.

## 2.3 Machine learning algorithms

This section explains the machine learning training algorithms that we utilize in this research. These algorithms include classical algorithms such as Random Forest (RF) and XGBoost (XGB), as well as deep learning algorithms such as Long Short-Term Memory (LSTM) and Gated Recurrent Unit (GRU).

The description of the four algorithms used in this research is provided below.

- **Random Forest (RF)**: Random decision forest (RF) is an ensemble learning method that builds multiple decision trees during training and can be used for regression and classification tasks. The main goal of RF is to combine the output (predictions) of these trees to improve the robustness and performance of the model. In RF, each tree is trained on a randomly split subset of data, and the results of all trees are combined at the end using averaging for regression and majority voting for classification to generate the final result. The RF improves the generalizations and reduces the overfitting (IBM, 2024; Wikipedia Contributors, 2024b; Günay, 2021).
- **XGBoost (XGB)**: eXtreme Gradient Boosting (XGB) is a machine learning model that uses ensemble learning and is suitable for regression and classification tasks (Jafari et al., 2025). XGB implements gradient-boosted decision trees designed for speed and performance. The XGB combines the prediction of multiple models (e.g. decision trees) iteratively to build a predictive model. It incrementally adds weak models (trees) to the ensemble, with each new model emphasizing more on the data points that were previously misclassified. Also, XGB prevents overfitting by regularization (Vidhya, 2018; NVIDIA, 2024).
- **Long Short-Term Memory (LSTM)**: LSTM is a special neural network structure that is designed to remember the relevant information for long periods when dealing with sequences like time series data, text, or speech. Unlike conventional recurrent neural networks that often confuse or “forget” details in long sequences, LSTMs use a set of “gates” to determine which pieces of information should be kept, updated, or removed from the network over time. This approach ensures that critical events at the start of the sequence can still be used when making predictions much later. The main strength of LSTMs is that they separate their "cell state" (which handles long-term memory) from a short- term hidden state, helping them avoid issues like exploding gradients or vanishing. The design choice of LSTMs helps them succeed in applications like predicting future trends in a time series, predicting words in a sequence, and translating between languages (Hochreiter and Schmidhuber, 1997; Gers et al., 2000).
- **Gated Recurrent Unit (GRU)**: GRUs are specialized types of recurrent neural networks designed to handle sequential information like time series, text, and audio by keeping track of important information over several time steps. They achieve this through using "gates" which act like regulators that decide how much new input to keep and how much old data to discard. At each point in a sequence, the GRU decides what new information should be added to the memory and what old information should be deleted. This design helps GRUs remember key details for a long time and discard irrelevant ones. It also helps GRUs avoid issues seen in traditional RNNs, such as "forgetting" significant parts of the sequence or having difficulty in training because of exploding or vanishing gradients (Goodfellow et al., 2016; Chung et al., 2014).

We utilize RF because it is effective for complex relationships and high-dimensional datasets (Undy, 2024). Since our datasets are multivariate, RF can capture the relationships between different features. RF also reduces overfitting by averaging the results of multiple decision trees. Besides, RF is scalable and robust enough to handle outliers (Undy, 2024).

We utilize XGB because it is good at capturing non-linear relationships. Identifying non-linear relationships is important for time series datasets, where patterns may not always be linear. It applies regularization to avoid overfitting. XGB is also scalable and able to handle outliers (Undy, 2024). Both RF and XGB can handle multiple input variables (Undy, 2024).

We use LSTM and GRU because these models are popular models used for time series forecasting (Goodfellow et al., 2016; Hochreiter and Schmidhuber, 1997; Chung et al., 2014).

### 2.4 Concept drift

In machine learning, when the distribution of the target concept changes over time, there is a "concept drift" (Nishida and Yamauchi, 2007; Widmer and Kubat, 1996; Gama et al., 2014). Concept drift and performance degradation are closely related phenomena in machine learning systems. The presence of one implies the existence of the other (Bayram et al., 2022). Concept drift can render models trained on earlier data obsolete, leading to a decline in performance (Lyu et al., 2021a). Reoccurring concept drift is a type of concept drift that refers to the reappearance of concepts over time (Zliobaite, 2010). This research targets reoccurring concept drifts.

### 2.5 Evaluation metrics

This section explains the metrics we used to evaluate our approach.

**Performance measurement metric**: Our datasets are time series data with numeric target values. We use 1) Mean Squared Error (MSE) (Wikipedia Contributors, 2024a) and 2) Mean Absolute Error (MAE) as measurement metrics to evaluate the performance of the machine learning models. MSE computes the average squared distance between the actual value and the predicted values. The MSE is given by Equation 1. MAE computes how far the predictions are from the actual values on average (Willmott and Matsuura, 2005). The MAE is given by Equation 2.

$$\text{MSE} = \frac{1}{n}\sum_{i=1}^{n}(y_i - \hat{y}_i)^2 \tag{1}$$

$$\text{MAE} = \frac{1}{n}\sum_{i=1}^{n}|y_i - \hat{y}_i| \tag{2}$$

**Statistical comparison metric**: We aim to compare our approach's performance with the baseline approaches' performance. Therefore, we use a non-parametric pairwise statistical test to compare the results. Specifically, we use the Mann-Whitney U statistical test to compare the results of our approach with those of the baseline approach. We choose Mann-Whitney U because it is a non-parametric test that does not assume that the data is normally distributed (which is not true in the case of our data). Also, this test is suitable when having a small sample size, which is true in our case. Finally, this test is appropriate when the samples are independent (each row of results is independent of the other rows of results).

The threshold we picked to separate significant and insignificant differences is 0.05 for the p-value (Lyu et al., 2021a). The p-values greater or equal to 0.05 show an insignificant difference between the results of the two compared approaches. The p-values less than 0.05 show a significant difference between the results of the two approaches.

**Estimated maintenance time and cost**: we use operation time as metrics for model maintenance costs. These times are calculated using Equation 3, where "operation" refers to the processes of training, prediction, forecasting, and similarity measurement.

$$\text{Operation time} = \text{operation end time} - \text{operation start time} \tag{3}$$

In addition, we calculate the financial cost of our approach and compare it to the financial cost of baseline approaches. We run the codes using a Macbook Pro version 2021 with an Apple M1 Pro chip, eight cores (six performance and two efficiency), and 16GB memory, and MacOS Ventura 13.1. To measure the financial cost of running the code, we measure the cost of computations using the equivalent AWS SageMaker instance. The equivalent AWS Sagemaker instance to our MacBook Pro specifications is ml.m5.large (Amazon Web Services, 2024a). The hourly rate of this instance is $0.115. To calculate how much our computations cost, we use the Equation 4.

$$\text{Financial Cost} = \text{Hourly Rate} \times \frac{Minutes}{60} \tag{4}$$

## 3 Related works

In this section, we first briefly review related prior work and highlight their limitations.

### 3.1 Impacts of concept on model performance

Concept drift is a common challenge that companies face when deploying their models. For instance, (Lyu et al., 2021a) shows that concept drift exists in the AIOps datasets like the Google cluster trace dataset (Wilkes, 2020) and the Backblaze disk stats dataset (Backblaze, 2024). Concept drift can lead to model performance degradation because the models have not seen the new data distributions and are not trained for the new patterns. Therefore, model maintenance plays a critical role in handling concept drifts in production.

There are different types of drift that can occur in the data. Zliobaite (2010) discusses four different types of drift that may occur in the dataset, as shown in Figure 2. In sudden drift, the distribution of data shifts instantly at a certain point in time. In gradual drift, the data distribution changes slowly from one concept to another, with overlapping phases. In incremental drift, the data distribution changes step-by-step over time in a continuous manner. In reoccurring contexts (reoccurring concept drift), previously observed data distributions reappear periodically or sporadically over time. In this study, we examine the reocurrence of data distribution and seasonalities in time series datasets over time. Thus, we target the reoccurring concept drifts (existence of reoccurring contexts) of the time series datasets (UbiOps Contributors, 2024). By reocurring drift, we refer to situations where the same contexts reappear over time, causing repeated changes in the data.

In subsections 3.2, we explain the model maintenance and drift mitigation approaches proposed by the literature and their shortcomings.

### 3.2 Model maintenance and Concept drift mitigation

In this section, we explain the proposed model maintenance approaches and the efforts made to compare different approaches. We also explain the studies that tried to mitigate drift.

Lyu et al. (2021a) show a method to maintain and update models that are deployed in production to reduce the risk of concept drift. They presented that periodically updating models can mitigate the model performance degradation and concept drift issues, outperforming the performance of the stationary model (the initial model never gets updated in the stationary model approach). However, periodically retraining the model is more costly than stationary models (Lyu et al., 2021a).

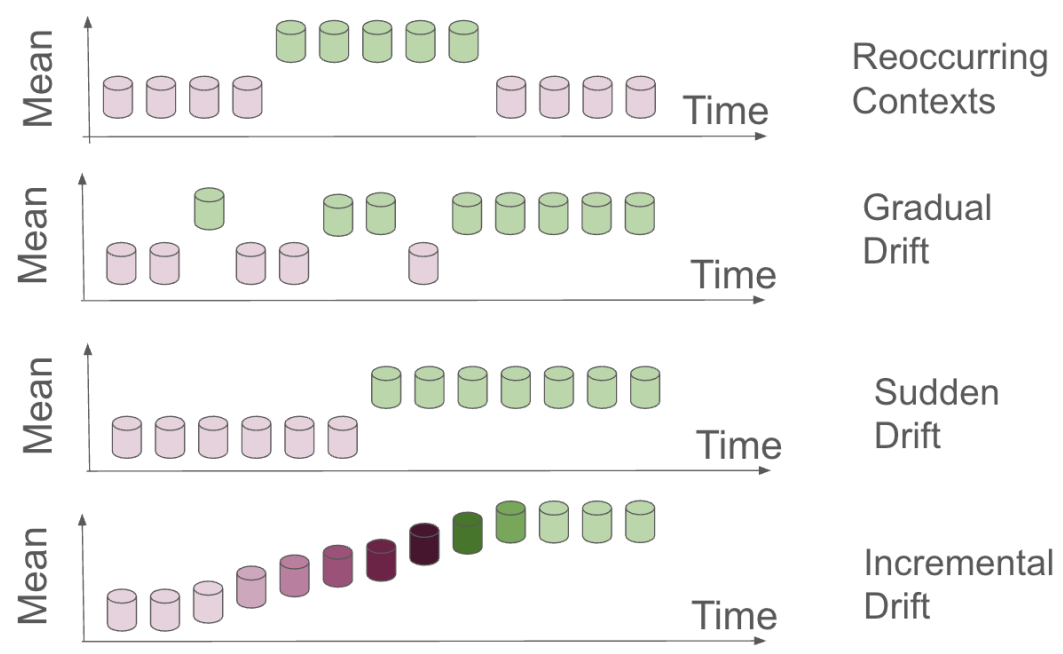

Fig. 2: Four different types of drift (Zliobaite, 2010)

Poenaru-Olaru et al. (2023) propose an automated maintenance pipeline for AIOps models that detects concept drift and selects the most suitable model retraining method based on the type of drift detected. (Poenaru-Olaru et al., 2022) show that class imbalance, which is unavoidable in operational data, negatively impacts the accuracy of the concept drift detectors. Although they limited the model training to the number of drifts in the dataset, they did not reuse the existing models.

Previous works also mention two other types of model maintenance and retraining techniques: “full history” and “sliding window” (Lyu et al., 2021b; Poenaru-Olaru et al., 2024). The full history approach continuously combines the training dataset with the recent data and retrains the model. The sliding window method retrains the model with only the most recent data as the training data while discarding old samples.

Although both full history and sliding window approaches achieve good results, based on the specific scenario, one approach may perform better than the other (Lyu et al., 2021a; Poenaru-Olaru et al., 2023). Sometimes, the drift occurs seasonally. For example, an increase in the usage of servers may cause more job failures. Server usage might rise during certain times, leading to an increase in job failures. In this example, the full history method is better because the full history is trained on data that captures multiple periods and distributions. However, the sliding window method may not have seen all the seasonal distributions (Poenaru-Olaru et al., 2023).

Lyu et al. (2024) conducted an empirical study of five different model update approaches for supervised learning, evaluating them in terms of stability, update costs, and performance. Their results demonstrate that concept drift-guided retraining, periodic retraining, online learning, and time-based model ensembles offer superior performance and greater stability compared to stationary models. However, their findings also indicate that, despite better performance, online learning and time-based ensembles increase model testing time. Consequently, their use in AIOps applications is limited, as these applications require immediate inferences.

To mitigate the impact of concept drift on model performance, some research papers propose methods for detecting (Gama et al., 2004; Harel et al., 2014; Nishida and Yamauchi, 2007; Wang et al., 2013) and handling concept drift (Brzezinski and Stefanowski, 2013; Cano and Krawczyk, 2020; Dongre and Malik, 2014; Gama et al., 2014; Hoens et al., 2012; Minku et al., 2009; Minku and Yao, 2011; Street and Kim, 2001; Wang et al., 2003). Poenaru-Olaru et al. (2024) propose a concept drift mitigation technique for anomaly detection systems called "Informed model update". They suggest that instead of blindly updating anomaly detection models, it is better to update them whenever a change is detected in the data. This method helps to avoid unnecessary retraining.

Mallick et al. (2022) propose a flexible, adaptable, and scalable data drift mitigation solution (Matchmaker) for deploying large-scale systems. Some of the differences between our approach and the Matchmaker are 1) Matchmaker focuses on large-scale systems, while our approach can handle any time series datasets with seasonality and reoccurring patterns, 2) Matchmaker identifies similar batches for each test point, while our approach uses the overall distribution similarity for each data segment, reducing the inference time, 3) Match-Maker reactively chooses the correct model when a new data instance arrives, which causes additional latency introduced by comparing the new instance with existing data, whereas our approach proactively forecasts the distribution in the future and proactively chooses the model. Therefore, MatchMaker has more overhead compared to our approach. Table 1 shows the other differences between Matchmaker and SimReuse. Our model reuse approach clusters the data based on the seasonality and similar data distributions, while Matchmaker does not consider the distribution of data while clustering it. Our approach focuses on the reocurring and seasonal drifts, while it is not the case in Matchmaker approach. Also, our approach relies on offline computations.

Table 1: Differences between Matchmaker and SimReuse

| Features | SimReuse | Matchmaker |
|---|---|---|
| Clusters the data based on seasonal/recurrent patterns or similar distributions | ✓ | ✗ |
| Proactive approach | ✓ | ✗ |
| Data clusters represent similar sets of data distributions | ✓ | ✗ |
| Reduces the frequency of periodically updating models | ✓ | ✗ |
| Focuses on time series datasets with reoccurring and seasonal drifts | ✓ | ✗ |
| Relies on offline computations | ✓ | ✗ |

In summary, the aforementioned prior works have the following shortcomings:

- The current model maintenance techniques, such as periodically updating the model, are not efficient enough in terms of time and cost of training and testing. Some of the existing approaches (e.g. online learning) are also model-dependent.
- To the best of our knowledge, none of the previous studies have considered the reocurring and seasonal patterns in data distributions to leverage existing models and avoid unnecessary retraining.
- Existing approaches detect drift when the new data arrives, but we are predicting the future distribution and proactively choosing the ML model. Some prior works use the existing data in their approach, but it is not realistic in the production environment because we do not have future data .

In this study, we aim to propose a new model maintenance approach to address model maintenance in case of reoccurring concept drift. We also propose an improved MLOps pipeline that developers can use when creating their own MLOps pipeline.

## 4 Preliminary study

In this section, we conduct a preliminary study to analyze the distribution patterns in the datasets, aiming to discover whether recurrent concept drifts and repeating seasonal patterns exist in the time series datasets.

Figure 3 presents the overview of the preliminary study design. First, we choose the datasets and preprocess them. Next, we find the right window size to divide each dataset into smaller segments. Then, we use forecasting methods to predict what the data distribution will look like in the future in the production environment, and we use similarity measurement methods to find which past data window it is most similar to. Finally, we study the similarities between the forecasted data segments and the earlier ones. The process of dataset selection and design of the preliminary study is explained in more detail below.

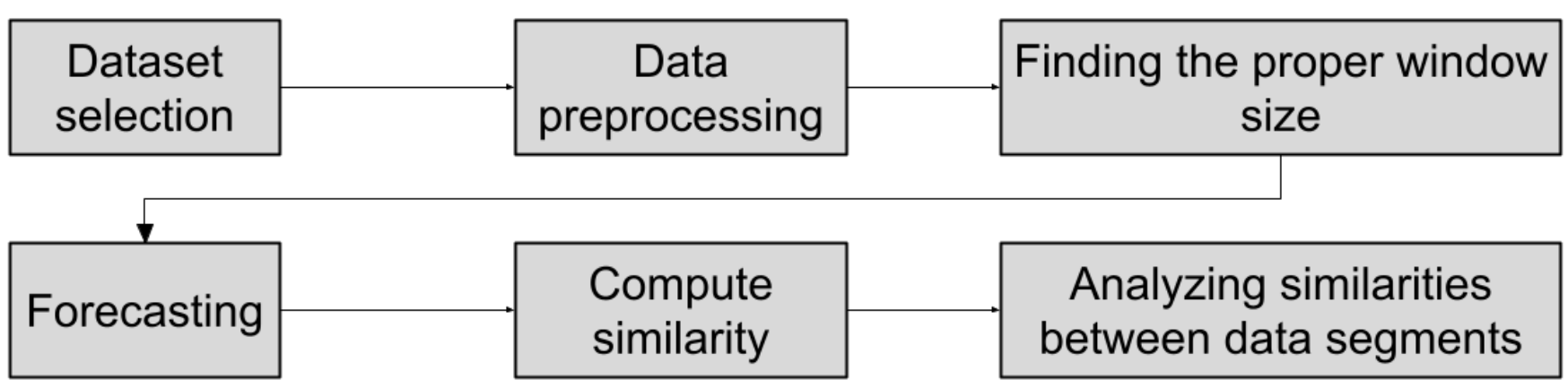

Fig. 3: Overview of the preliminary study

### 4.1 Datasets

To select our datasets, we applied the following criteria: the datasets should 1) be real, not synthetic data; 2) be used in other research papers, 3) span at least one year, and 4) be multivariate time series. Five public multivariate time series datasets are selected for this study based on the criteria. The datasets are described below. Also, more information about the start date, end date, number of total samples, number of features, number of samples per day, and time interval between the samples is provided in Table 2. We chose these five datasets because they are real datasets, and they are widely used in other research papers like (Zliobaite, 2013; Mallick et al., 2022; Das et al., 2023; Jin et al., 2023; Patil et al., 2024). It is worth mentioning that Electricity Zurich, Oil temperature, and Weather datasets are selected among the eight real-time series datasets that the Dart Python library provides (Herzen et al., 2022).

- **NSW Electricity** (Neto and Canuto, 2021): The electricity dataset includes data gathered from the New South Wales Electricity Market in Australia, where electricity prices are not fixed. Instead, prices fluctuate based on demand and supply over the last five minutes. The task is to determine the electricity price of the New South Wales (NSW) city in Australia based on other features (Neto and Canuto, 2021). The electricity dataset is influenced by concept drift because of shifting consumption patterns, unforeseen events, and seasonal variations. It has been used in over 40 concept drift experiments (Zliobaite, 2013). Prior studies have used the electricity dataset to evaluate their proposed approaches that target datasets with concept drift (Mallick et al., 2022; Neto and Canuto, 2021; Abbasi et al., 2021). The features of this dataset are also normalized between zero and one.
- **Zurich Electricity** (Herzen et al., 2022; Zürich, 2022, 2025): The Zurich Electricity dataset includes electricity usage by businesses and services (medium voltage), as well as by households and small-to-medium enterprises (SMEs) (low voltage), in the city of (Zürich, 2022). The electricity consumption data is integrated with hourly weather measurements from three stations in (Zürich, 2025). The original weather data, recorded hourly, is resampled to a 15-minute frequency, with any missing values interpolated before merging with the electricity consumption data (Unit8co, 2024). The task here is to forecast the electricity consumption of Households and SMEs. More description about the features of this dataset is provided in (Unit8co, 2024).
- **Oil temperature** (Herzen et al., 2022; Zhou, 2024; Zhou et al., 2021): This dataset includes the data of one electricity transformer at one station (Unit8co, 2024). The task here is to forecast the oil temperature. More description about the features of the dataset in provided in (Unit8co, 2024).
- **Weather** (Herzen et al., 2022; Max Planck Institute for Biogeochemistry, 2024; Zeng et al., 2023): This dataset, recorded every 10 minutes, includes

Table 2: Additional details of the datasets

| Dataset | Start date | End date | No. of total samples | No. of features | No. of samples per day | Time interval between the samples | Target value |
|---|---|---|---|---|---|---|---|
| Electricity NSW | 1996-06-3 | 1998-12-05 | 4,4016 | 4 | 48 | 30 min | NSW electricity price |
| Zurich Electricity | 2020-01-01 | 2020-12-31 | 35,040 | 9 | 96 | 15 min | Zurich electricity consumption |
| Oil temperature | 2016-07-01 | 2018-06-25 | 17,400 | 6 | 24 | 60 min | Oil temperature |
| Weather | 2020-01-01 | 2020-12-31 | 52,704 | 20 | 144 | 10 min | Weather temperature |
| NIFTY-50 | 2008-1-3 | 2020-12-30 | 3,198 | 10 | 1 | 24 hour | Close price |

21 indicators of the weather in Germany like humidity and air temperature(Unit8co, 2024). The task here is to forecast the weather temperature.

- **NIFTY-50 Stock Market** (Patil et al., 2024; Jain et al., 2018): This dataset includes the stock price history and the trading volumes of the fifty stocks in NIFTY-50 index from National Stock Exchange (NSE) India. The task here is to forecast the close price. Close price is the last agreed-upon value between sellers and buyers before the market closes for the day.

## 4.2 Design of the preliminary study

This section explains how the preliminary study is conducted.

**Data preprocessing**: To preprocess the data, all features of all datasets are normalized to values in the range of [0, 1] using the Min-Max scaler algorithm implemented by the scikit-learn library (Scikit-learn Developers, 2024), ensuring comparable feature contributions. Additionally, all dates are converted to the *datetime* format using the Pandas *datetime* function (The Pandas Development Team, 2024).

**Finding the proper window size**: We aim to train models on data windows with different data distributions and reuse the models in the future. One of the steps to achieve this goal is to find the proper window length to segment the data. To do so, we use the Autocorrelation Function (ACF) to find the reoccurring patterns in the data distribution.

*The Autocorrelation Function (ACF)* is a statistical tool that quantitatively measures the similarity between the time series and the lagged versions of itself over varying time intervals (lags). By comparing the original series with lagged versions, the correlation at each lag indicates how much past values resemble current values. Strong correlations at particular lags often indicate the presence of repeating behavior or cycles within the data (Box et al., 2015). For example, if a time series follows a repeating pattern with a particular period $p$, the autocorrelation will be higher at a lag of $p$. In other words, the values of the series at time $t$ are more similar to the values of the series at time $t$ - $p$ when there exists a repeating pattern every $p$ steps.

The main application of the ACF in time series analysis is detecting seasonalities. If the ACF detects high similarity at specific intervals, it represents the presence of recurring and consistent behaviors at those time intervals. For instance, when analyzing daily data, strong correspondence at intervals of 14 days often represents a biweekly pattern. This information is critical for building predictive models like resource planning, sales forecasting, and so on. Repeated rhythms significantly impact the outcome of predictive models (Hyndman and Athanasopoulos, 2018). In this research, we utilize ACF to assess if there are reoccurring patterns in the evaluated time series datasets.

*Our application of ACF:* There are a few steps to use ACF to find the recurring patterns in our studied datasets: Step 1) We first compute the daily average of the data over time and then apply the ACF on the daily average of the data. If the dataset has only one value per day, we apply the ACF on the data itself. Step 2) We analyze a varying range of lags ranging from 1 to half of the length of the dataset. We choose half of the length of the dataset to cover enough data to detect most of the possible seasonality lengths and leave enough data as the test data. Step 3) We go through the correlation values and find local maxima, which are points where the autocorrelation forms the "peak". We sort the autocorrelation peaks decrementally based on their values. The first peak lag (e.g. lag $l$) with the highest correlation value represents seasonality in every $l$ step. Step 4) We apply the following filters to pick the most suitable lag. From the sorted peak lags, we pick the first lag greater or equal to 14 days and less than nine months. We decided to pick this range to have enough data for training and testing the models. If there is no peak lag in this range, we pick the peak lag shorter than 14 days.

**Forecasting:** In our proposed approach, we compare the similarities between a new time window's data with previous time windows to identify reoccurring patterns. However, in practical scenarios, the entire data in the new time window may not be available (i.e., unseen) when we make predictions for instances in the window. For instance, if it is the beginning of February, we do not have the distribution pattern for the entire month of February. Therefore, we need to use a forecasting method to forecast the distribution for February.

We use the following forecasting approaches to approximate the distribution of the new unseen data in production. The approach selection reason is explained in detail in the approach description.

– *Approach 1 - Exponential Smoothing (ES)*: ES is a forecasting algorithm for the time series dataset that uses weighted averages of past observations to forecast future values, where weights diminish exponentially as the data ages. This algorithm assigns greater importance to recent observations while exponentially decreasing the importance of older data points (Simplilearn, 2023). This technique is suitable for datasets with seasonal patterns.
– *Approach 2 - Similarity Assumption (SA)*: Mallick et al. (2022) mentioned that models trained on training data with distribution A can be accurate for any test data with distribution B where B is similar to A. Similar to the idea presented in prior work (Brzezinski and Stefanowski, 2013; Tahmasbi et al., 2020), we assume that the data distribution in the upcoming window W will follow the data distribution from the past window. Throughout this paper, the output of forecasting methods is called forecasted data.

**Compute similarity:** We use two similarity measurement methods: 1)the Wasserstein distance (WD) and 2) total variation distance (TVD) to compute the similarity of each data window to its preceding windows. Since we don't know the distribution of the immediate future data where we will use the time series models, we use these similarity measures to compare the forecasted distributions in the immediate-future window with those of previous data windows. The forecasted data are produced by forecasting approaches 1 (ES) and 2 (SA). Then we compute the ratio of similar segments to their non-adjacent previous data segments. For computing the similarity ratio, we use Equation 7 presented below.

We decided to run our experiments with two different similarity measurement metrics to examine the impact that a choice of similarity measurement method has on the results of our study. Both similarity measurement techniques are applied on the ES and SA forecasting results. The description of the two methods is provided below.

– *Wasserstein distance (WD)*: Wasserstein distance (WD) or Earth Mover's Distance (EMD) computes the distance between two probability distributions over a given metric space. WD quantifies the least amount of effort needed to change one distribution into another. The amount of effort is measured by the distribution mass that needs to be moved times the mean distance that has to be moved (Wikipedia Contributors, 2024d). The reason for choosing WD is that it concerns where the probability mass is located and how far it must be moved (Amazon Web Services, 2024b). Probability mass means the amount of probability assigned to specific locations in the space under the distribution.
  As seen in Equation (5), $W_p(\mu,\nu)$ distance computes the minimum "cost" of transporting mass between the distributions $\mu$ and $\nu$. In this equation, $\mu$ and $\nu$ are probability distributions on a space $\mathcal{X}$, $d(x,y)$ is a distance function that computes how far $x$ is from $y$, and $p$ is a positive real number that depicts the type of Wasserstein distance. The set $\Pi(\mu,\nu)$ contains all couplings (or joint distributions) whose marginals are $\mu$ and $\nu$, and the expression $\inf_{\pi\in\Pi(\mu,\nu)}$ denotes that we choose the coupling $\pi$ that minimizes the integral of $d(x,y)^p$ with respect to $\pi$. In the end, the integral $\int_{\mathcal{X}\times\mathcal{X}} d(x,y)^p \, d\pi(x,y)$ sums (integrates) the cost of moving one unit of mass from $x$ to $y$. Raising this integral to the power of $1/p$ completes the description of the Wasserstein distance (Aubin-Frankowski and Rudi, 2024).

$$W_p(\mu,\nu) = \left( \inf_{\pi\in\Pi(\mu,\nu)} \int_{\mathcal{X}\times\mathcal{X}} d(x,y)^p \, d\pi(x,y) \right)^{\frac{1}{p}} \tag{5}$$

- *Total variation distance (TVD)*: The total variation distance (TVD) measures the difference between two probability distributions. Let $P$ and $Q$ be two probability distributions over the same sample space. $P(x)$ and $Q(x)$ are the probability of the event x happening based on the distributions P and Q, respectively (Reddit Contributors, 2024). In this case, TVD is the maximum difference in probabilities that these distributions assign to the same events (Levin and Peres, 2017; Wikipedia Contributors, 2024c). Interpretation of TVD is straightforward, making it a suitable choice for our study.

$$d_{\mathrm{TV}}(P, Q) = \frac{1}{2} \sum_{x} |P(x) - Q(x)| \tag{6}$$

$$\text{Similarity ratio} = \frac{\text{Number of similar non adjacent segments}}{\text{Total number of segments}} \tag{7}$$

**Analysis method:** The steps we follow to do our preliminary study are: The process begins by normalizing the dataset columns. Next, the daily average of the target variable is calculated. Then, ACF is applied to check if seasonality exists in the datasets. If seasonality is detected, we will pick the segment length that is most proper. Once the proper segment length is established, forecasting methods are applied to predict the target values for the upcoming data segment.

Then, the distribution similarity between the forecasted segment of data and previous segments of data is computed using the WD and TVD metrics. This way, we identify the most similar preceding segment of data.

## 4.3 Preliminary study results

This section provides the results achieved from the preliminary study.

### *4.3.1 Finding the proper window size*

We present the result achieved by ACF in this section. After applying ACF, we realized that there are seasonalities in the studied datasets. Table 3 shows the seasonal length found for each dataset. The NSW Electricity and the Zurich Electricity datasets have biweekly seasonal patterns. The Oil temperature dataset has a weekly seasonal pattern. In other words, we found weekly or biweekly seasonal patterns in 60% of our studied datasets. The Weather dataset and NIFTY-50 datasets have 15 days and 258 days of seasonal patterns, respectively.

In the next step, we divide the datasets into segments with the achieved seasonality length and then compute the similarity between segments as described below.

**Similarity measurement**: We measured the similarity between each segment with two similarity measurement approaches of Wasserstein distance and

Table 3: ACF seasonality results

| Dataset | Seasonality length (days) |
|---|---|
| NSW Electricity | 14 |
| Zurich Electricity | 14 |
| Oil temperature | 7 |
| Weather | 15 |
| NIFTY-50 | 258 |

TVD. Table 4 shows the ratio of similar segments for each forecasting method for each dataset. Similar segments have a similar data distributions.

Table 4: Percentage of similar segment ratio. In the names of columns two to five, A - B means the forecasting method is A and the similarity measurement method is B.

| Dataset | SA - Wass | SA - TVD | ES - Wass | ES - TVD |
|---|---|---|---|---|
| NSW Electricity | 74% | **82%** | 82% | 63% |
| Zurich Electricity | 46% | **69%** | 50% | 61% |
| Oil temperature | 79% | **94%** | 86% | 67% |
| Weather | 58% | **70%** | 62% | 50% |
| NIFTY50 | 41% | **66%** | 50% | 58% |

As Table 4 shows, there are high similarities (more than 41% for all datasets) between the segments and their non-adjacent preceding segments. Oil temperature has the highest similarity among its segments. NIFTY-50 shows the lowest similarity among its segments. In 70% of the cases, the TVD similarity measurement method achieved a higher ratio over the Wass method. Results achieved in Table 4 show that the SA forecasting method mixed with the TVD similarity measurement method achieves the highest similarity ratio for all datasets.

**Summary of preliminary study results**

- There are considerable re-occurrences of similar data distributions over the time span of our studied time series datasets.
- We found weekly or biweekly seasonal patterns in 60% of our studied datasets.
- All of the studied datasets show more than 41% similarities between the segments and their preceding segments, indicating the feasibility of our approach that leverages the similarity between anticipated distributions and previous distributions to reuse previously trained models.

## 5 Proposed solution (model reuse approach)

In the preliminary study, we found seasonal and reoccurring distribution patterns in time series datasets. Therefore, in this section, we aim to reuse our previously trained models in the future for similar data distributions. We propose our model maintenance approach, which is model-agnostic and requires less maintenance time and cost, to help MLOps engineers and companies save resources and time.

First, we propose our modified MLOps pipeline and its components and explain how it works in the production environment. Then, we explain and implement our model reuse approach in our SimReuse tool.

### 5.1 Proposed MLOps pipeline

The proposed MLOps pipeline is shown in Figure 4. Some components of our MLOps pipeline and their description below are adapted from existing work (Majidi et al., 2022; Amershi et al., 2019; Hepworth, 2022; Google Cloud, 2024). The subscript “New” identifies the components that we add to the existing MLOps pipeline. Each new component is explained in more detail in what follows:

**Distribution analysis$_{\mathbf{New}}$**: We follow a two-step process to analyze how the target column changes over time. First, we calculate the daily average of the target column to find any change in the target data distributions. Then, we create distribution and box plots to show how these daily averages are distributed over the year. After analyzing the data distribution and box plots over time, we call the best window length that separates the patterns as “W”.

**Model registry**: Like the MLOps pipeline, we store the trained models in the model registry.

**Forecasting$_{\mathbf{New}}$**: Our tool aims to work in real-time systems. Since we do not have the data distribution for future data, we need approaches to forecast the target values in the next window. We use two approaches, SA and ES, to forecast the target values in the next window.

**Compute Similarity$_{\mathbf{New}}$**: One essential contribution of this research is reducing the frequency of model training by reusing existing models. To achieve this, we propose utilizing the same model for several windows with data distributions similar to the train data. The initial step involves calculating the distribution similarity score between each window and its preceding windows. We use two methods, including 1) Wasserstein distance and 2) total variation distance, to compute the distribution similarity scores.

After computing the distribution similarity score between each window and its preceding windows, we identify the most similar window. This window is the one among the previous windows that is most similar to the inference data. The most similar window can be the previous window or one of the earlier windows.

**Model reuse$_{\mathbf{New}}$**: Training a machine learning model is costly and time-consuming. Therefore, in the model reuse component, we aim to reuse a model trained on a similar distribution to our current inference data instead of training a new model.

A new model is trained on the most recent data segment if it is the most similar data segment. Otherwise, we reuse the model trained on the most similar data segment. In other words, models trained on data segments with distribution "$X$" are reused for data segments with distribution "$Y$", provided that $Y$ is similar to $X$, as determined by the similarity scores from the similarity measurement step. This model reuse component is explained in detail in section 5.2.

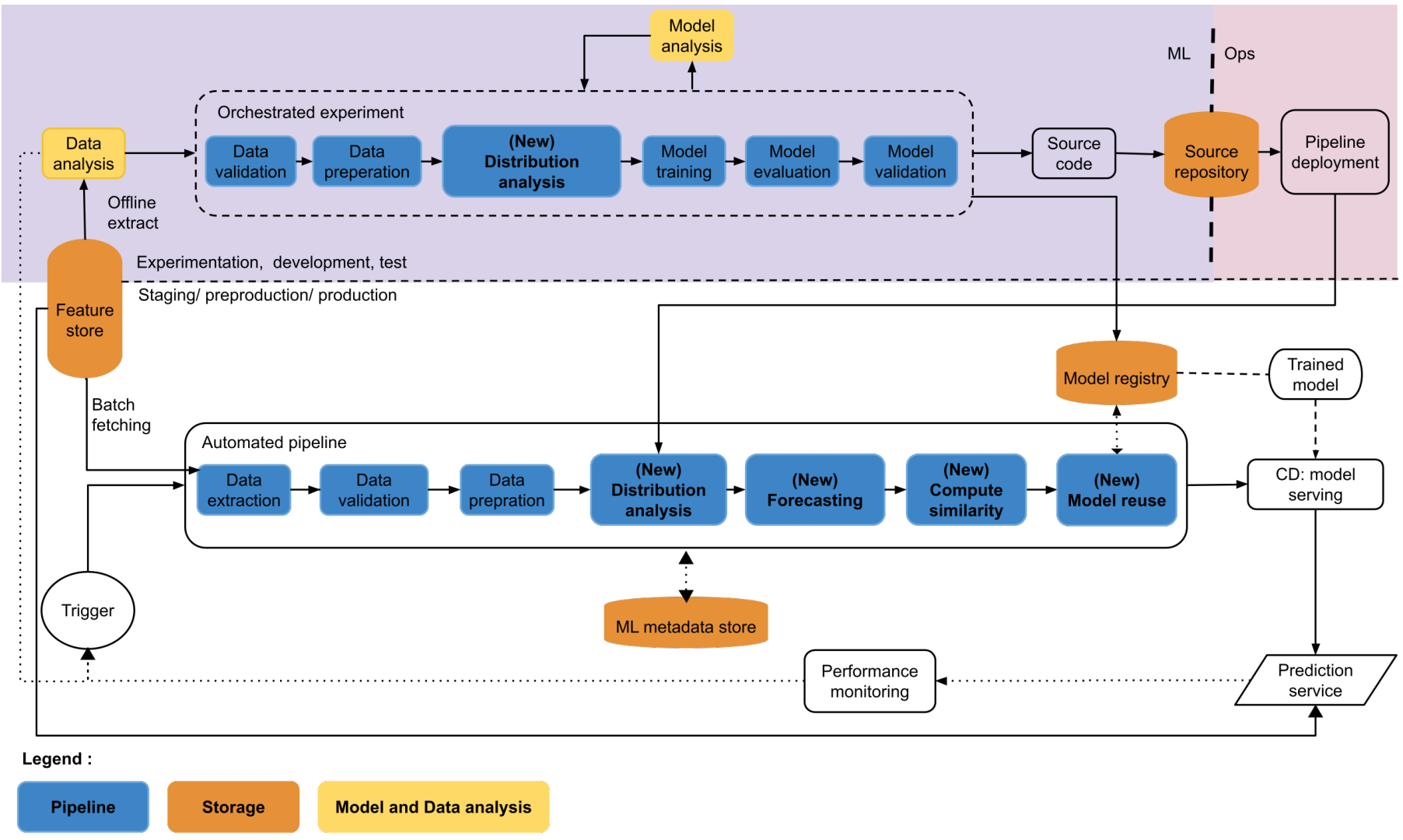

Fig. 4: Proposed improved MLOps pipeline

### 5.2 Model reuse

In this section, we explain our model reuse approach. There is a "model reuse" component in the proposed MLOps pipeline in Figure 4. Figure 5 shows an example of the model reuse approach. Also, Figure 6 breaks down this component using a flowchart to explain it in more detail.

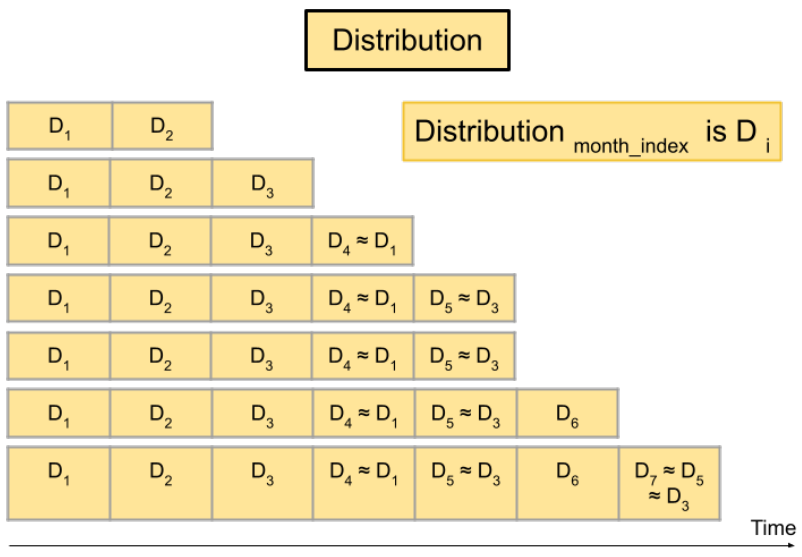

(a) An example of data distribution in each time window

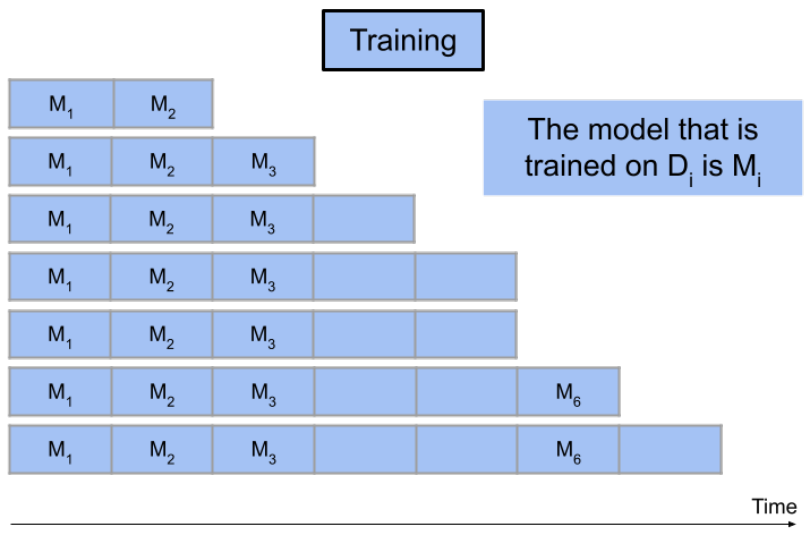

(b) Continuing the example: Models that are trained on data of each time window

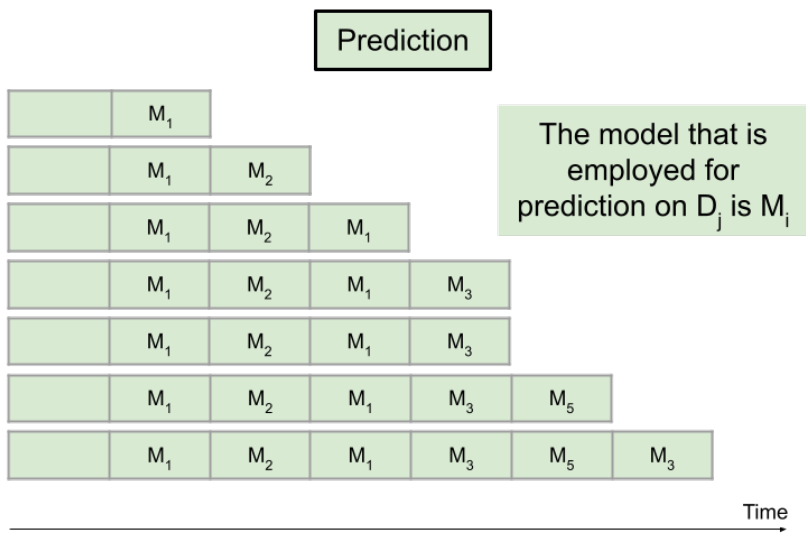

(c) Continuing the example: The model used for prediction in each time window

Fig. 5: An example of the model reuse approach

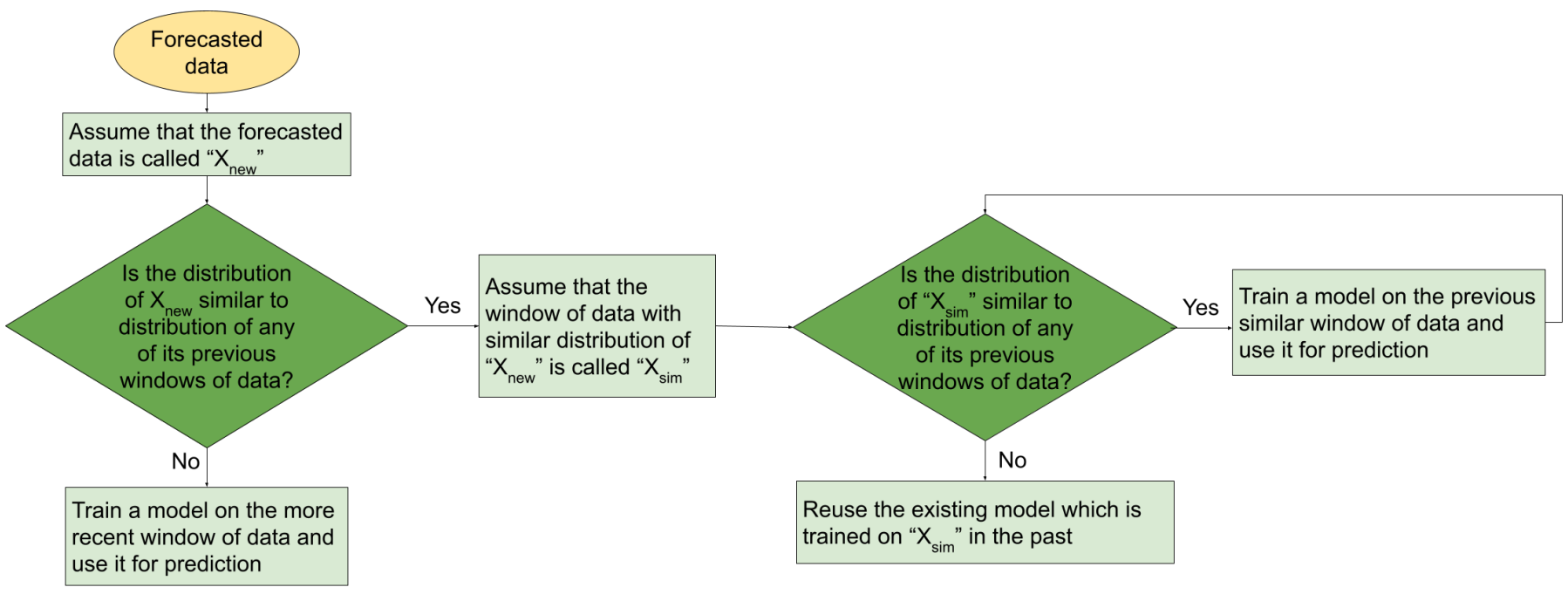

Fig. 6: Model reuse flowchart

The algorithm 1 explains the methodology for model reuse step by step. The step-by-step explanation of the algorithm 1 is given below:

– (Line 1) The inputs of the algorithm are 1) the *dataset*, 2) the size of the window (*window-size* "W"), 3) the *targetcolumn*, 4) a *windows_ similarity_ dictionary*

**Algorithm 1** Model Reuse Algorithm

```
Input: dataset, window_size, target_column, windows_similarity_dictionary, forecasting_approach
Initialize: saved_models ← [ ]                                   ▷ store the models in a list
for i ← window_size to len(dataset) step window_size do         ▷ loop through the dataset in windows
    train ← dataset[i− window_size : i]                          ▷ assign train data
    test ← dataset[i : i+ window_size]                           ▷ assign test data
    y_train ← train[target_column]
    X_train ← train.drop(columns=[target_column])
    y_test ← test[target_column]
    X_test ← test.drop(columns=[target_column])
    if len(test) < window_size then
        break                                   ▷ break if the test size is less than window_size
    end if
    if forecasting_approach == "ES" then
        window_index ← round(i/window_size) − 1    ▷ calculate window index for ES approach
    else if forecasting_approach == "SA" then
        window_index ← round(i/window_size)        ▷ calculate window index for SA approach
    end if
    if window_index ∈ windows_similarity_dictionary then   ▷ check if window index is in similarity dictionary
        similar_window_index ← windows_similarity_dictionary[window_index]   ▷ find similar window index
        if similar_window_index ∈ windows_similarity_dictionary then   ▷ check if there is a chain in similarities
            if forecasting_approach == "ES" then
                previous_model_i ← (window_index + 1) × window_size
            else if forecasting_approach == "SA" then
                previous_model_i ← window_index × window_size
            end if
            test2 ← dataset[previous_model_i : previous_model_i + window_size]
            y_test2 ← test2[target_column]
            X_test2 ← test2.drop(columns=[target_column])
            model ← train_model(X_test2, y_test2)   ▷ train model M on the previous test data
            saved_models[similar_window_index] ← model   ▷ replace the stored model with the new one
            append model to saved_models                 ▷ save the new model
            y_pred ← model.predict(X_test)
            MSE ← mean_squared_error(y_test, y_pred)     ▷ calculate MSE
            MAE ← mean_absolute_error(y_test, y_pred)    ▷ calculate MAE
        else
            model ← saved_models[similar_window_index]   ▷ retrieve the existing model
            append model to saved_models                 ▷ save the reused model
            y_pred ← model.predict(X_test)
            MSE ← mean_squared_error(y_test, y_pred)     ▷ calculate MSE
            MAE ← mean_absolute_error(y_test, y_pred)    ▷ calculate MAE
        end if
    else
        model ← train_model(X_train, y_train)   ▷ train a new model on the most recent window
        y_pred ← model.predict(X_test)
        MSE ← mean_squared_error(y_test, y_pred)         ▷ calculate MSE
        MAE ← mean_absolute_error(y_test, y_pred)        ▷ calculate MAE
        append model to saved_models                     ▷ save the new model
    end if
end for
```

where the keys are windows indices, and the values are the corresponding similar windows to the keys, and 5) *forecastingapproach*.
- (Line 2) We store all the models in a list called *saved_ models*.
- (Line 3 to line 49) We go through the *dataset* in a for loop.
- (Line 4 to line 12) *train* and *test* sets are assigned. We exit the for loop if the test data length is shorter than the *windowsize*.
- (Line 13 to line 17) We assign the *window_ index* based on the *forecasting-approach*.
- (Line 18 to line 19) If the *window_ index* is among the values in the *windows_ similarity_ dictionary*, assign the key to the *similar_ window_ index*.
- (Line 20 to line 30) If the *similar_ window_ index* is among the values in the *windows_ similarity_ dictionary*, it means that we have a chain of similarities. If we have a chain of similarities, we train a model (B) on the data of the *similar_ window_ index*.
- (Line 31 to line 32) We store the *model* and use it for prediction on the test data.
- (Line 33 to line 34) We measure the mean squared error and mean absolute error.
- (Line 35 to line 41) If we do not have a chain of similarities, we simply get the saved model and use it to make the prediction. Then, we measure the mean squared error and mean absolute error.
- (Line 42 to line 49) If the *similar_ window_ index* is not among the values in the *windows_ similarity_ dictionary*, we train a model on the most recent window of data and use it to make predictions. Then, we measure the mean squared error and mean absolute error. Finally we store the *model* for further reuse in the future.

It is worth mentioning that (Lyu et al., 2021a) mentioned that the most frequently used model maintenance and concept drift adaptation method is model retraining. However, the retraining model takes time and is costly. Some companies with large models and massive datasets may be unable to afford it. Besides, the Stationary model's performance may degrade over time. Thus, our model reuse preserves the model performance while keeping the maintenance time and cost low.

## 6 Evaluation

We proposed our approach in Section 5. In this section, we evaluate the effectiveness of our method. In this section, we first explain our analysis approach; then, we present the results that we have achieved.

### 6.1 Evaluation context

This section explains the steps we took to evaluate our approach.

**Baseline selection process** To evaluate the effectiveness of our approach, we aim to compare our approach with state-of-art baselines. To find the baselines, we follow the steps below:

- We search for related papers on Google Scholar and Engineering Village (Compendex, Inspect databases).
- We search for the following query to find the related papers: ("Software engineering" OR aiops OR mlops) AND ((data OR concept OR model) AND (drift OR shift)) AND (mitigat* OR retrain OR maintenance OR maintain* OR detect OR detection). We aim to include all keywords and concepts related to model maintenance and MLOps that might be mentioned in other research works (e.g. AIOps).
- We filter the papers based on their publication year, discarding those published before 2018.
- To select the baselines, we follow the criteria mentioned below:
  - Our approach is model-agnostic, which is compatible with all models and time series datasets. Consequently, one of our baseline selection criteria is its model-agnostic nature.
  - We discard the approaches that are made for specific algorithms, systems, purposes, or use cases.
  - The baseline approaches must have an immediate inference time because time is important in some applications (e.g. AIOps applications) (Lyu et al., 2024).
  - The baseline approaches must have been used in other research papers.

After following the steps above, we identify three baseline approaches that are widely used in the related research papers (Lyu et al., 2021a; Poenaru-Olaru et al., 2024). The baseline approaches are 1) the stationary model, 2) the periodically updated model, and 3) the informed model retraining. The description of the baseline methods and their training process is provided below. Also, we include a random prediction model as our last baseline.

**1) Periodically updated model**: In the periodically updated model approach, a model is trained on the most recent time window of data (e.g. last month) and tested on the next time window of data (test data). Note that the sliding window of the training data and test data move forward over time. Besides, we do not use all the historical data to update the model. Instead, we use only the latest sliding window. This is because using all the data is hard to scale, especially when the model needs frequent updates. Moreover, using all the past data may not necessarily improve the model's performance as much as using just the new data. Figure 7b shows the periodically updated model approach (Lyu et al., 2021a).

We provide examples of the prediction phase of periodically updated in Figure 9. Figure 9 shows how the periodically updated model makes predictions in each time window. In Figure 9, each box is a time window, and $\mathrm{M_k}$ is the model trained on the time window "k".

*Periodically updated model training process*: During the model retraining process, we employ four distinct algorithms, RF, XGB, LSTM, and GRU to minimize the impact of algorithm choice on the results. We periodically update the models with a wide range of segment lengths, including five days, 15 days, 31 days, 45 days, 60 days, 75 days, and 90 days for the NSW Electricity, Zurich Electricity, Oil temperature, Weather datasets, and store their MSE and MAE. For the NIFTY-50 dataset, we consider bigger segment lengths because we have more years of data available. The segment lengths we used for periodically retraining the NIFTY-50 dataset are 50 days, 150 days, 310 days, 450 days, 600 days, 750 days, and 900 days. Having a wide range of segment lengths allows us to have full coverage of different possible segment lengths. Then, we periodically update the model with the segment length achieved the least MSE.

**2) Stationary model**: In the Stationary model approach, a model is trained on a first time window (e.g., the first month) and never gets updated. Figure 7a shows the stationary model approach (Lyu et al., 2021a).

We provide examples of the prediction phase of the stationary models in Figure 9 and 8. Figure 8 depicts the stationary model's prediction phase, indicating the model used for predictions in each time window. In Figures 8, $M_1$ is the model trained on the training data.

*Stationary model training process*: We use four different algorithms, RF, XGB, LSTM, and GRU to lessen the impact of training methods on our results. Then, we train a stationary model on the selected segment length by the model reuse approach.

**3) Random baseline**: We generate random predictions within the range of minimum and maximum values of the inference data's target column. Then, we evaluate the predicted result.

**4) Informed retraining (Poenaru-Olaru et al., 2024)**: In the informed retraining approach, the model is retrained whenever drift is detected in the recent test data segment (Poenaru-Olaru et al., 2024). To detect the drift, we followed the approach used by Lyu et al. (2021a). Following the (Poenaru-Olaru et al., 2024) approach, concept drift is detected by training a model on one time period and testing it on the next, then using a two-proportion Z-test to compare the model's prediction error rates between the training and testing periods. If the statistical test shows a significant difference (p-value $< 0.05$), it indicates that concept drift has occurred.If the drift is detected for a segment of test data, the model is retrained on the previous segment of data. If no drift is detected, the last trained model is used until a drift occurs in a new test data segment. Figure 7c shows the informed model retraining approach. As shown in Figure 7c, drift is detected in the third test data segment. Thus, a model retraining message is sent to update the model. Also, Figure 10 shows an example of the prediction phase of the informed retraining approach. Additionally, we measure the ratio of drifted segments of data based on the Equation 8.

$$\text{Drift ratio} = \frac{\text{Number of concept drifted segments}}{\text{Total number of segments}} \tag{8}$$

*Informed retraining process*: We first detect the drifts using the method mentioned in (Lyu et al., 2021a). Secondly, we retrain the model on the last segment of the data if the drift occurs. During the model retraining process, we employ four distinct algorithms, RF, XGB, LSTM, and GRU, to minimize the impact of algorithm choice on the results.

Table 5 shows the ratio of concept-drifted segments. As seen in Table 5, all of our studied datasets have drifted segments. The Oil temperature has the lowest drifted segment ratio with 31% drift, and the Zurich electricity dataset has the highest drifted segments with 76% drift.

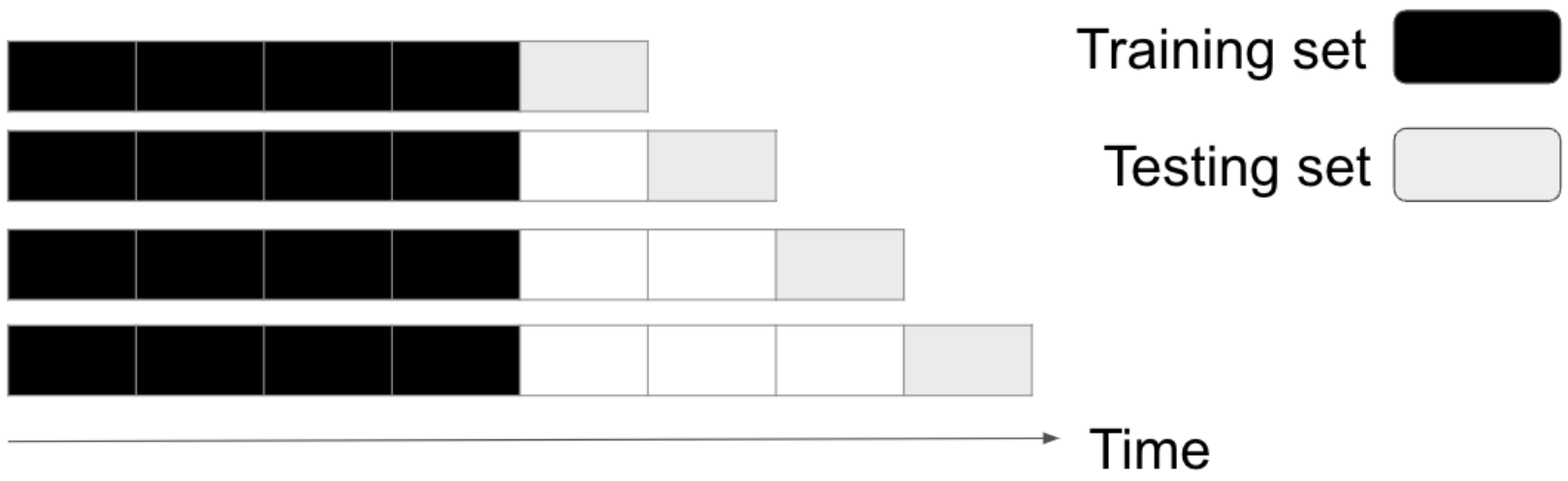

(a) Stationary model approach (Lyu et al., 2021a)

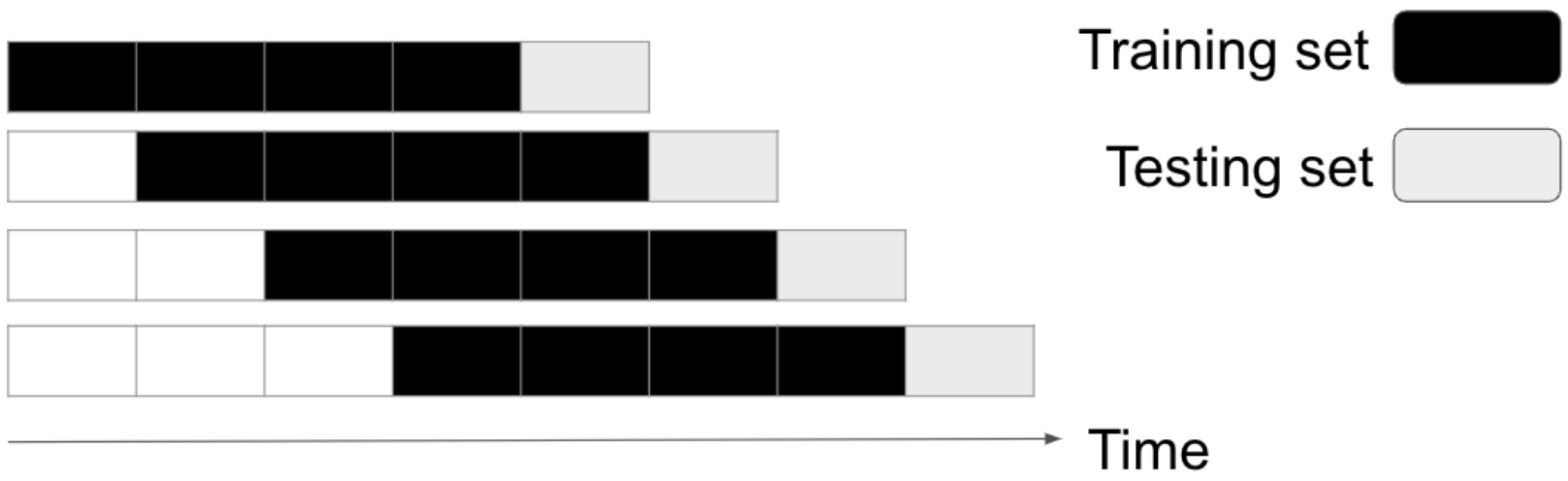

(b) Periodically updated model approach (Lyu et al., 2021a)

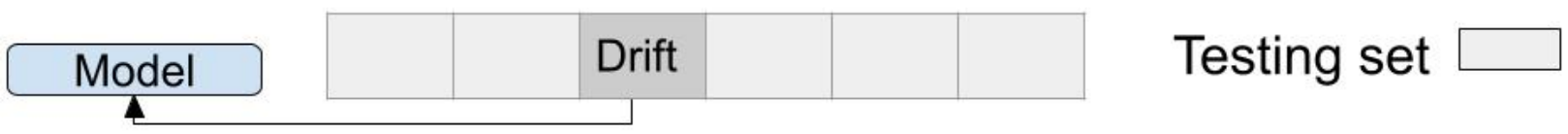

(c) Informed retraining approach (Poenaru-Olaru et al., 2024)

Table 5: Percentage of concept drifted segments

| Dataset | Percentage of concept drifted segments |
|---|---|
| NSW Electricity | 50% |
| Zurich Electricity | 76% |
| Oil temperature | 31% |
| Weather | 69% |
| NIFTY-50 | 36% |

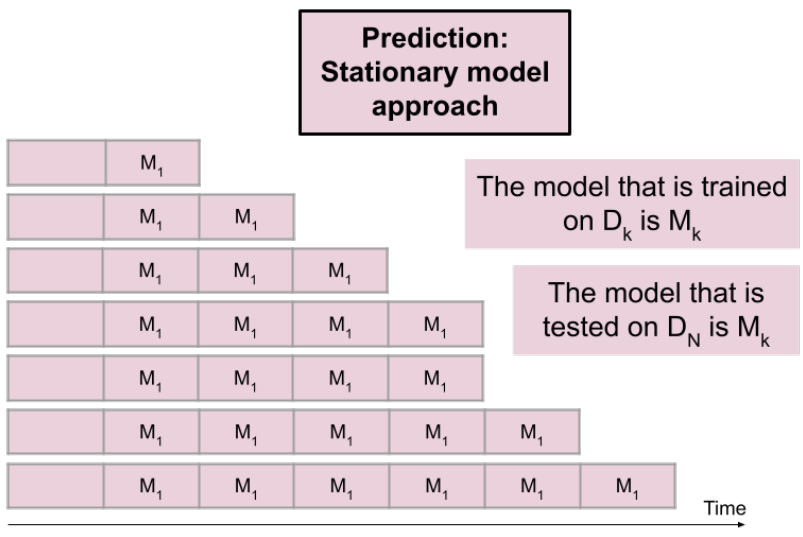

Fig. 8: An example of the prediction phase of the stationary model approach. $D_K$ is the distribution of segment K. $M_K$ is the model that is trained on $D_K$.

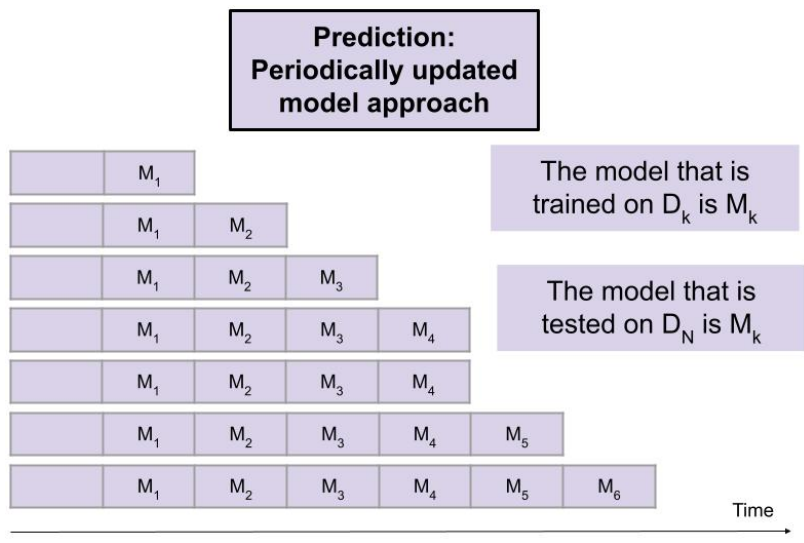

Fig. 9: An example of the prediction phase of the periodically updated model approach. $D_K$ is the distribution of segment K. $M_K$ is the model that is trained on $D_K$.

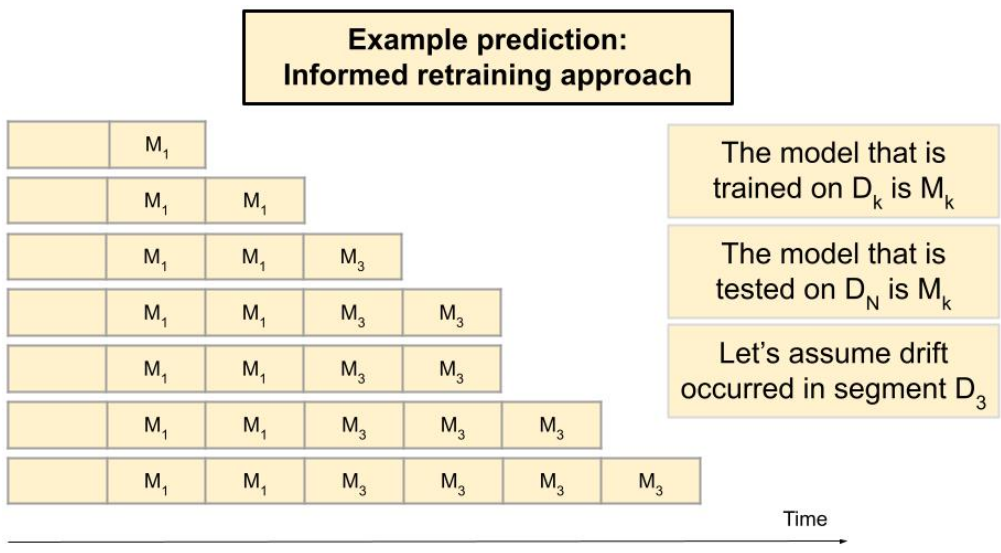

Fig. 10: An example of the prediction phase of the informed retraining approach. $D_K$ is the distribution of segment K. $M_K$ is the model that is trained on $D_K$.

### 6.2 RQ1: How does SimReuse perform regarding ML performance metrics?

We have proposed a new model maintenance approach and developed our SimReuse tool so far. In this section, we aim to evaluate the performance of our proposed method.

**Analysis method:** To evaluate SimReuse with the baselines, we follow the steps below: We developed and applied five distinct modeling approaches - periodically updated models, stationary models, informed retraining, random models, and model reuse strategies — across all five datasets. After implementing these approaches, we conducted a performance comparison to evaluate the effectiveness of each learning algorithm on different datasets.

In the proposed solution in section 5, we develop SimReuse to reduce the number of unnecessary retrainings. In RQ1, we aim to evaluate its performance and compare it with the baselines.

**Results:** Tables 6 and 7 show the MSE and MAE results of the model reuse approach using LSTM and RF as the learning algorithms in four different settings. They also compare the results of different baseline approaches. Please note that the performance results of the GRU and the XGB learning algorithms are presented in the appendix.

We run the model reuse approach for each ML learning algorithm for each dataset with four different configurations. These four configurations are different combinations of results from two similarity metrics (WD and TVD) and two forecasting methods (SA and ES). The reported results are only for the time windows for which we utilized the "model reuse" approach for their inference. The best two performances in each row are highlighted in bold text.

Table 7 represents the MSE and MAE results of the model reuse approach employing RF as the learning algorithm under four different configurations. They also compare the results of different baseline approaches.

As seen in the result Tables 6 and 7, the periodically model update approach achieves the lowest MSE and MAE values among the baselines and model reuse approach in most of the situations. However, there are cases highlighted by ** in the result tables where the model reuse approach achieves the lowest error compared to other baselines, even to the periodical retraining baseline. According to the results presented in Tables 6 and 7, 50%, 60%, 60% and 70% of the configurations, respectively, achieved either the lowest or second-lowest error using the model reuse approach after periodical retraining. Furthermore, the results of the Tables 6 and 7 show that the forecasting method and similarity metric do not have a significant effect on the MSE and MAE of the model reuse approach.

**Comparison of the results**: We use the Mann-Whitney U statistical test to compare the results of different approaches. We compare the results of the model reuse approach with all of the other approaches. Table 8 shows the comparison between the different approaches. To compare the results of different approaches, we concatenated the MSE and MAE results of different learning algorithms. Then we applied the Mann-Whitney U statistical test on the results. **As seen in Table 8, the comparison between the periodically updated model and the model reuse approach achieves the highest p-value greater than 0.05, which shows an insignificant difference between the results. Therefore, this result shows that we can use the model reuse approach instead of periodical retraining approaches.** This way, we can save time, cost, and resources. Furthermore, the comparison between the informed retraining and the model reuse approach gives a high p-value (above 0.05), meaning there is also no significant difference between their results. Moreover, stationary and random baselines have the smallest p-value (much smaller than 0.05) when compared with the model reuse approach. This result shows a greater difference between these two baselines and the model reuse approach.

Table 6: Performance comparison of different model update approaches (learning algorithm: LSTM). The first two lowest error values are highlighted in each row. ** sign highlights the rows where the model reuse error is lower than the periodical error.

(a) Performance measurement metric: MSE

| Dataset | Similarity metric | Forecast approach | Stationary model MSE | periodically updated model MSE | model reuse MSE | Informed retraining MSE | Random model MSE |
|---|---|---|---|---|---|---|---|
| NSW Electricity | W | SA | $19 \times 10^{-4}$ | $\mathbf{14 \times 10^{-4}}$ | $\mathbf{12 \times 10^{-4}}$ | $19 \times 10^{-4}$ | $34 \times 10^{-3}$ |
| NSW Electricity | TVD | SA | $19 \times 10^{-4}$ | $\mathbf{13 \times 10^{-4}}$ | $\mathbf{11 \times 10^{-4}}$ | $20 \times 10^{-4}$ | $34 \times 10^{-3}$ |
| NSW Electricity | W | ES | $19 \times 10^{-4}$ | $\mathbf{14 \times 10^{-4}}$ | $\mathbf{11 \times 10^{-4}}$ | $17 \times 10^{-4}$ | $34 \times 10^{-3}$ |
| NSW Electricity | TVD | ES | $19 \times 10^{-4}$ | $\mathbf{15 \times 10^{-4}}$ | $\mathbf{14 \times 10^{-4}}$ | $15 \times 10^{-4}$ | $34 \times 10^{-3}$ |
| Zurich Electricity | W | SA | $13 \times 10^{-3}$ | $\mathbf{11 \times 10^{-3}}$ | $\mathbf{12 \times 10^{-3}}$ | $14 \times 10^{-3}$ | $59 \times 10^{-3}$ |
| Zurich Electricity | TVD | SA | $13 \times 10^{-3}$ | $\mathbf{11 \times 10^{-3}}$ | $19 \times 10^{-3}$ | $\mathbf{11 \times 10^{-3}}$ | $59 \times 10^{-3}$ |
| Zurich Electricity | W | ES | $13 \times 10^{-3}$ | $\mathbf{11 \times 10^{-3}}$ | $13 \times 10^{-3}$ | $\mathbf{11 \times 10^{-3}}$ | $59 \times 10^{-3}$ |
| Zurich Electricity | TVD | ES | $13 \times 10^{-3}$ | $\mathbf{12 \times 10^{-3}}$ | $14 \times 10^{-3}$ | $\mathbf{10 \times 10^{-3}}$ | $59 \times 10^{-3}$ |
| Oil temperature ** | W | SA | $57 \times 10^{-3}$ | $\mathbf{63 \times 10^{-4}}$ | $\mathbf{51 \times 10^{-4}}$ | $91 \times 10^{-4}$ | $10 \times 10^{-3}$ |
| Oil temperature | TVD | SA | $57 \times 10^{-3}$ | $\mathbf{62 \times 10^{-4}}$ | $31 \times 10^{-3}$ | $\mathbf{84 \times 10^{-4}}$ | $10 \times 10^{-3}$ |
| Oil temperature ** | W | ES | $57 \times 10^{-3}$ | $\mathbf{63 \times 10^{-3}}$ | $\mathbf{46 \times 10^{-4}}$ | $10 \times 10^{-3}$ | $10 \times 10^{-3}$ |
| Oil temperature | TVD | ES | $57 \times 10^{-3}$ | $\mathbf{63 \times 10^{-4}}$ | $\mathbf{10 \times 10^{-3}}$ | $11 \times 10^{-3}$ | $10 \times 10^{-3}$ |
| Weather | W | SA | $12 \times 10^{-4}$ | $\mathbf{16 \times 10^{-5}}$ | $\mathbf{17 \times 10^{-5}}$ | $22 \times 10^{-5}$ | $37 \times 10^{-3}$ |
| Weather | TVD | SA | $12 \times 10^{-4}$ | $\mathbf{17 \times 10^{-5}}$ | $31 \times 10^{-5}$ | $\mathbf{21 \times 10^{-5}}$ | $37 \times 10^{-3}$ |
| Weather | W | ES | $12 \times 10^{-4}$ | $\mathbf{15 \times 10^{-5}}$ | $45 \times 10^{-5}$ | $\mathbf{19 \times 10^{-5}}$ | $37 \times 10^{-3}$ |
| Weather | TVD | ES | $12 \times 10^{-4}$ | $\mathbf{16 \times 10^{-5}}$ | $47 \times 10^{-5}$ | $\mathbf{18 \times 10^{-5}}$ | $37 \times 10^{-3}$ |
| NIFTY-50 | W | SA | $58 \times 10^{-4}$ | $\mathbf{19 \times 10^{-4}}$ | $37 \times 10^{-4}$ | $\mathbf{35 \times 10^{-4}}$ | $30 \times 10^{-3}$ |
| NIFTY-50 | TVD | SA | $58 \times 10^{-4}$ | $\mathbf{26 \times 10^{-4}}$ | $58 \times 10^{-4}$ | $\mathbf{53 \times 10^{-4}}$ | $30 \times 10^{-3}$ |
| NIFTY-50 | W | ES | $58 \times 10^{-4}$ | $\mathbf{20 \times 10^{-4}}$ | $67 \times 10^{-4}$ | $\mathbf{52 \times 10^{-4}}$ | $30 \times 10^{-3}$ |
| NIFTY-50 ** | TVD | ES | $58 \times 10^{-4}$ | $\mathbf{24 \times 10^{-4}}$ | $\mathbf{22 \times 10^{-4}}$ | $49 \times 10^{-4}$ | $30 \times 10^{-3}$ |

(b) Performance measurement metric: MAE

| Dataset | Similarity metric | Forecast approach | Stationary model MAE | periodically updated model MAE | model reuse MAE | Informed retraining MAE | Random model MAE |
|---|---|---|---|---|---|---|---|
| NSW Electricity | W | SA | $29 \times 10^{-3}$ | $\mathbf{16 \times 10^{-3}}$ | $\mathbf{16 \times 10^{-3}}$ | $22 \times 10^{-3}$ | $10 \times 10^{-2}$ |
| NSW Electricity | TVD | SA | $29 \times 10^{-3}$ | $\mathbf{16 \times 10^{-3}}$ | $\mathbf{17 \times 10^{-3}}$ | $22 \times 10^{-3}$ | $10 \times 10^{-2}$ |
| NSW Electricity | W | ES | $29 \times 10^{-3}$ | $\mathbf{16 \times 10^{-3}}$ | $\mathbf{16 \times 10^{-3}}$ | $19 \times 10^{-3}$ | $10 \times 10^{-2}$ |
| NSW Electricity | TVD | ES | $29 \times 10^{-3}$ | $\mathbf{16 \times 10^{-3}}$ | $\mathbf{19 \times 10^{-3}}$ | $\mathbf{19 \times 10^{-3}}$ | $10 \times 10^{-2}$ |
| Zurich Electricity | W | SA | $91 \times 10^{-3}$ | $\mathbf{82 \times 10^{-3}}$ | $\mathbf{85 \times 10^{-3}}$ | $91 \times 10^{-3}$ | $19 \times 10^{-2}$ |
| Zurich Electricity | TVD | SA | $91 \times 10^{-3}$ | $\mathbf{83 \times 10^{-3}}$ | $10 \times 10^{-2}$ | $\mathbf{83 \times 10^{-3}}$ | $19 \times 10^{-2}$ |
| Zurich Electricity | W | ES | $91 \times 10^{-3}$ | $\mathbf{82 \times 10^{-3}}$ | $88 \times 10^{-3}$ | $\mathbf{83 \times 10^{-3}}$ | $19 \times 10^{-2}$ |
| Zurich Electricity | TVD | ES | $91 \times 10^{-3}$ | $\mathbf{84 \times 10^{-3}}$ | $93 \times 10^{-3}$ | $\mathbf{78 \times 10^{-3}}$ | $19 \times 10^{-2}$ |
| Oil temperature | W | SA | $20 \times 10^{-2}$ | $\mathbf{61 \times 10^{-3}}$ | $\mathbf{54 \times 10^{-3}}$ | $72 \times 10^{-3}$ | $74 \times 10^{-3}$ |
| Oil temperature | TVD | SA | $20 \times 10^{-2}$ | $\mathbf{63 \times 10^{-3}}$ | $11 \times 10^{-2}$ | $\mathbf{72 \times 10^{-3}}$ | $80 \times 10^{-3}$ |
| Oil temperature ** | W | ES | $20 \times 10^{-2}$ | $\mathbf{61 \times 10^{-3}}$ | $\mathbf{51 \times 10^{-3}}$ | $80 \times 10^{-3}$ | $80 \times 10^{-3}$ |
| Oil temperature | TVD | ES | $20 \times 10^{-2}$ | $\mathbf{62 \times 10^{-3}}$ | $\mathbf{76 \times 10^{-3}}$ | $84 \times 10^{-3}$ | $80 \times 10^{-3}$ |
| Weather | W | SA | $24 \times 10^{-3}$ | $\mathbf{95 \times 10^{-4}}$ | $\mathbf{96 \times 10^{-4}}$ | $11 \times 10^{-3}$ | $15 \times 10^{-2}$ |
| Weather | TVD | SA | $24 \times 10^{-3}$ | $\mathbf{91 \times 10^{-4}}$ | $11 \times 10^{-3}$ | $\mathbf{96 \times 10^{-4}}$ | $15 \times 10^{-2}$ |
| Weather | W | ES | $24 \times 10^{-3}$ | $\mathbf{91 \times 10^{-4}}$ | $13 \times 10^{-3}$ | $\mathbf{99 \times 10^{-4}}$ | $15 \times 10^{-2}$ |
| Weather | TVD | ES | $24 \times 10^{-3}$ | $\mathbf{94 \times 10^{-4}}$ | $13 \times 10^{-3}$ | $\mathbf{10 \times 10^{-3}}$ | $15 \times 10^{-2}$ |
| NIFTY-50 | W | SA | $70 \times 10^{-3}$ | $\mathbf{32 \times 10^{-3}}$ | $\mathbf{40 \times 10^{-3}}$ | $44 \times 10^{-3}$ | $10 \times 10^{-2}$ |
| NIFTY-50 | TVD | SA | $70 \times 10^{-3}$ | $\mathbf{35 \times 10^{-3}}$ | $74 \times 10^{-3}$ | $\mathbf{60 \times 10^{-3}}$ | $10 \times 10^{-2}$ |
| NIFTY-50 | W | ES | $70 \times 10^{-3}$ | $\mathbf{33 \times 10^{-3}}$ | $\mathbf{60 \times 10^{-3}}$ | $66 \times 10^{-3}$ | $10 \times 10^{-2}$ |
| NIFTY-50 | TVD | ES | $70 \times 10^{-3}$ | $\mathbf{36 \times 10^{-3}}$ | $\mathbf{30 \times 10^{-3}}$ | $61 \times 10^{-3}$ | $10 \times 10^{-2}$ |

Table 7: Performance comparison of different model update approaches (learning algorithm: RF). The first two lowest error values are highlighted in each row. ** sign highlights the rows where the model reuse error is lower than the periodical error.

(a) Performance measurement metric: MSE

| Dataset | Similarity metric | Forecast approach | Stationary model MSE | periodically updated model MSE | model reuse MSE | Informed re-training MSE | Random model MSE |
|---|---|---|---|---|---|---|---|
| NSW Electricity ** | W | SA | $17 \times 10^{-4}$ | $\mathbf{97 \times 10^{-5}}$ | $\mathbf{86 \times 10^{-5}}$ | $12 \times 10^{-4}$ | $34 \times 10^{-3}$ |
| NSW Electricity ** | TVD | SA | $17 \times 10^{-4}$ | $\mathbf{97 \times 10^{-5}}$ | $\mathbf{51 \times 10^{-5}}$ | $16 \times 10^{-4}$ | $34 \times 10^{-3}$ |
| NSW Electricity ** | W | ES | $17 \times 10^{-4}$ | $\mathbf{97 \times 10^{-5}}$ | $\mathbf{38 \times 10^{-5}}$ | $93 \times 10^{-5}$ | $34 \times 10^{-3}$ |
| NSW Electricity ** | TVD | ES | $17 \times 10^{-4}$ | $\mathbf{97 \times 10^{-5}}$ | $\mathbf{97 \times 10^{-5}}$ | $83 \times 10^{-5}$ | $34 \times 10^{-3}$ |
| Zurich Electricity ** | W | SA | $95 \times 10^{-4}$ | $\mathbf{63 \times 10^{-4}}$ | $\mathbf{58 \times 10^{-4}}$ | $94 \times 10^{-4}$ | $60 \times 10^{-3}$ |
| Zurich Electricity | TVD | SA | $95 \times 10^{-4}$ | $\mathbf{63 \times 10^{-4}}$ | $77 \times 10^{-4}$ | $\mathbf{74 \times 10^{-4}}$ | $60 \times 10^{-3}$ |
| Zurich Electricity | W | ES | $95 \times 10^{-4}$ | $\mathbf{63 \times 10^{-4}}$ | $70 \times 10^{-4}$ | $\mathbf{67 \times 10^{-4}}$ | $60 \times 10^{-3}$ |
| Zurich Electricity | TVD | ES | $95 \times 10^{-4}$ | $\mathbf{63 \times 10^{-4}}$ | $61 \times 10^{-4}$ | $\mathbf{58 \times 10^{-4}}$ | $60 \times 10^{-3}$ |
| Oil temperature ** | W | SA | $12 \times 10^{-2}$ | $\mathbf{55 \times 10^{-4}}$ | $\mathbf{42 \times 10^{-4}}$ | $21 \times 10^{-3}$ | $14 \times 10^{-3}$ |
| Oil temperature | TVD | SA | $12 \times 10^{-2}$ | $\mathbf{55 \times 10^{-4}}$ | $62 \times 10^{-3}$ | $\mathbf{14 \times 10^{-3}}$ | $14 \times 10^{-3}$ |
| Oil temperature ** | W | ES | $12 \times 10^{-2}$ | $\mathbf{55 \times 10^{-4}}$ | $\mathbf{53 \times 10^{-4}}$ | $10 \times 10^{-3}$ | $14 \times 10^{-3}$ |
| Oil temperature | TVD | ES | $12 \times 10^{-2}$ | $\mathbf{55 \times 10^{-4}}$ | $82 \times 10^{-4}$ | $\mathbf{56 \times 10^{-4}}$ | $14 \times 10^{-3}$ |
| Weather ** | W | SA | $12 \times 10^{-3}$ | $\mathbf{23 \times 10^{-5}}$ | $\mathbf{12 \times 10^{-5}}$ | $71 \times 10^{-5}$ | $37 \times 10^{-3}$ |
| Weather | TVD | SA | $12 \times 10^{-3}$ | $\mathbf{23 \times 10^{-5}}$ | $10 \times 10^{-4}$ | $\mathbf{41 \times 10^{-5}}$ | $37 \times 10^{-3}$ |
| Weather | W | ES | $12 \times 10^{-3}$ | $\mathbf{23 \times 10^{-5}}$ | $21 \times 10^{-4}$ | $\mathbf{50 \times 10^{-5}}$ | $37 \times 10^{-3}$ |
| Weather | TVD | ES | $12 \times 10^{-3}$ | $\mathbf{23 \times 10^{-5}}$ | $28 \times 10^{-4}$ | $\mathbf{33 \times 10^{-5}}$ | $37 \times 10^{-3}$ |
| NIFTY-50 | W | SA | $33 \times 10^{-4}$ | $\mathbf{15 \times 10^{-4}}$ | $\mathbf{20 \times 10^{-4}}$ | $26 \times 10^{-4}$ | $30 \times 10^{-3}$ |
| NIFTY-50 | TVD | SA | $33 \times 10^{-4}$ | $\mathbf{15 \times 10^{-4}}$ | $\mathbf{33 \times 10^{-4}}$ | $62 \times 10^{-4}$ | $30 \times 10^{-3}$ |
| NIFTY-50 | W | ES | $33 \times 10^{-4}$ | $\mathbf{15 \times 10^{-4}}$ | $\mathbf{36 \times 10^{-4}}$ | $66 \times 10^{-4}$ | $30 \times 10^{-3}$ |
| NIFTY-50 | TVD | ES | $33 \times 10^{-4}$ | $\mathbf{15 \times 10^{-4}}$ | $\mathbf{24 \times 10^{-4}}$ | $45 \times 10^{-4}$ | $30 \times 10^{-3}$ |

(b) Performance measurement metric: MAE

| Dataset | Similarity metric | Forecast approach | Stationary model MAE | periodically updated model MAE | model reuse MAE | Informed re-training MAE | Random model MAE |
|---|---|---|---|---|---|---|---|
| NSW Electricity | W | SA | $25 \times 10^{-3}$ | $\mathbf{88 \times 10^{-4}}$ | $\mathbf{90 \times 10^{-4}}$ | $11 \times 10^{-3}$ | $10 \times 10^{-2}$ |
| NSW Electricity | TVD | SA | $25 \times 10^{-3}$ | $\mathbf{88 \times 10^{-4}}$ | $\mathbf{91 \times 10^{-4}}$ | $12 \times 10^{-3}$ | $10 \times 10^{-2}$ |
| NSW Electricity ** | W | ES | $25 \times 10^{-3}$ | $\mathbf{88 \times 10^{-4}}$ | $\mathbf{82 \times 10^{-4}}$ | $92 \times 10^{-4}$ | $10 \times 10^{-2}$ |
| NSW Electricity | TVD | ES | $25 \times 10^{-3}$ | $\mathbf{88 \times 10^{-4}}$ | $10 \times 10^{-3}$ | $\mathbf{86 \times 10^{-4}}$ | $10 \times 10^{-2}$ |
| Zurich Electricity ** | W | SA | $78 \times 10^{-3}$ | $\mathbf{56 \times 10^{-3}}$ | $\mathbf{47 \times 10^{-3}}$ | $72 \times 10^{-3}$ | $19 \times 10^{-2}$ |
| Zurich Electricity | TVD | SA | $78 \times 10^{-3}$ | $\mathbf{56 \times 10^{-3}}$ | $\mathbf{57 \times 10^{-3}}$ | $62 \times 10^{-3}$ | $19 \times 10^{-2}$ |
| Zurich Electricity ** | W | ES | $78 \times 10^{-3}$ | $\mathbf{56 \times 10^{-3}}$ | $\mathbf{56 \times 10^{-3}}$ | $58 \times 10^{-3}$ | $19 \times 10^{-2}$ |
| Zurich Electricity ** | TVD | ES | $78 \times 10^{-3}$ | $\mathbf{56 \times 10^{-3}}$ | $\mathbf{47 \times 10^{-3}}$ | $55 \times 10^{-3}$ | $19 \times 10^{-2}$ |
| Oil temperature ** | W | SA | $31 \times 10^{-2}$ | $\mathbf{58 \times 10^{-3}}$ | $\mathbf{44 \times 10^{-3}}$ | $99 \times 10^{-3}$ | $94 \times 10^{-3}$ |
| Oil temperature | TVD | SA | $31 \times 10^{-2}$ | $\mathbf{58 \times 10^{-3}}$ | $16 \times 10^{-2}$ | $\mathbf{84 \times 10^{-3}}$ | $94 \times 10^{-3}$ |
| Oil temperature ** | W | ES | $31 \times 10^{-2}$ | $\mathbf{58 \times 10^{-3}}$ | $\mathbf{47 \times 10^{-3}}$ | $72 \times 10^{-3}$ | $94 \times 10^{-3}$ |
| Oil temperature | TVD | ES | $31 \times 10^{-2}$ | $\mathbf{58 \times 10^{-3}}$ | $63 \times 10^{-3}$ | $\mathbf{60 \times 10^{-3}}$ | $94 \times 10^{-3}$ |
| Weather ** | W | SA | $62 \times 10^{-3}$ | $\mathbf{37 \times 10^{-4}}$ | $\mathbf{20 \times 10^{-4}}$ | $78 \times 10^{-4}$ | $15 \times 10^{-2}$ |
| Weather | TVD | SA | $62 \times 10^{-3}$ | $\mathbf{37 \times 10^{-4}}$ | $84 \times 10^{-4}$ | $\mathbf{52 \times 10^{-4}}$ | $15 \times 10^{-2}$ |
| Weather | W | ES | $62 \times 10^{-3}$ | $\mathbf{37 \times 10^{-4}}$ | $14 \times 10^{-3}$ | $\mathbf{59 \times 10^{-4}}$ | $15 \times 10^{-2}$ |
| Weather | TVD | ES | $62 \times 10^{-3}$ | $\mathbf{37 \times 10^{-4}}$ | $10 \times 10^{-3}$ | $\mathbf{48 \times 10^{-4}}$ | $15 \times 10^{-2}$ |
| NIFTY-50 | W | SA | $35 \times 10^{-3}$ | $\mathbf{17 \times 10^{-3}}$ | $\mathbf{23 \times 10^{-3}}$ | $30 \times 10^{-3}$ | $10 \times 10^{-2}$ |
| NIFTY-50 | TVD | SA | $35 \times 10^{-3}$ | $\mathbf{17 \times 10^{-3}}$ | $\mathbf{36 \times 10^{-3}}$ | $57 \times 10^{-3}$ | $10 \times 10^{-2}$ |
| NIFTY-50 | W | ES | $35 \times 10^{-3}$ | $\mathbf{17 \times 10^{-3}}$ | $\mathbf{32 \times 10^{-3}}$ | $69 \times 10^{-3}$ | $10 \times 10^{-2}$ |
| NIFTY-50 | TVD | ES | $35 \times 10^{-3}$ | $\mathbf{17 \times 10^{-3}}$ | $\mathbf{21 \times 10^{-3}}$ | $51 \times 10^{-3}$ | $10 \times 10^{-2}$ |

Table 8: Comparison of different approaches performance results with Mann-Whitney U. Periodical means Periodically updated model approach. Reuse means model reuse approach. The model update approaches with the most insignificant error value differences are highlighted in the table.

| Compared approaches | p-value | Significant or insignificant |
|---|---|---|
| **Reuse & Periodical** | **0.945** | **insignificant** |
| Reuse & Informed | 0.574 | insignificant |
| Reuse & Stationary | $3 \times 10^{-6}$ | significant |
| Reuse & Random | $6 \times 10^{-30}$ | significant |

**Summary of RQ1 results**

- Model reuse achieves the lowest or second lowest error after periodical retraining in 60% of the tested configurations with different datasets.
- The difference between the model reuse approach and the periodic model updating approach is not statistically significant. Both approaches perform better than the stationary and random approaches (with statistically significant differences).
- We encourage developers and MLOps engineers to utilize the model reuse approach or SimReuse tool in their projects as they achieve comparable results as the periodically updated model approach while being able to reduce the frequency of model updating.

### 6.3 RQ2: How does SimReuse perform regarding maintenance costs?

We have analyzed SimReuse's performance so far. In RQ2, we aim to evaluate ML models' maintenance time and financial costs and compare them with the time and costs of the baseline approaches. Thus, in this section, we report the savings of the model reuse approach regarding computation time and computation costs. Time is reported in "second" in this section.

**Analysis method**: To evaluate our approach's computation time and financial cost versus the baseline approaches, we run all the approaches, measure their computation cost and time, and compare them at the end. The most resource-intensive computations involve several tasks across the periodically updated model, stationary model, and model reuse approaches.

In the periodically updated model, essential resource-intensive tasks include determining the segment length that achieves the minimum MSE and periodically updating the model with this segment length. In the stationary model approach, training and applying the model to the inference data are costly processes. Model reuse requires resources for forecasting with the Exponential Smoothing (ES) approach, computing similarity metrics, and applying the model reuse approach using both the SA and ES forecasting methods.

Lastly, informed model updating requires time and resources to run drift detection code and retrain the model whenever the drift is detected.

It's worth mentioning that we have two methods for forecasting: ES and SA. Only the ES approach is costly and time-consuming. Choosing the proper approach depends on the developer in the development phase, and we do not always have to use the ES approach when running the pipeline.

The Table 9 shows the number of reduced model reuse approach retrainings compared to periodical retraining approach

Table 9: Model reuse reduced number of retrainings compared to periodical retraining

| Dataset | Similarity metric | Forecast approach | Count of reduced retrainings | Percentage of reduced retraining count |
|---|---|---|---|---|
| NSW Electricity | W | SA | 27 times | 42% |
| NSW Electricity | TVD | SA | 35 times | 55% |
| NSW Electricity | W | ES | 28 times | 44% |
| NSW Electricity | TVD | ES | 29 times | 46% |
| Zurich Electricity | W | SA | 7 times | 26% |
| Zurich Electricity | TVD | SA | 13 times | 50% |
| Zurich Electricity | W | ES | 9 times | 34% |
| Zurich Electricity | TVD | ES | 9 times | 34% |
| Oil temperature | W | SA | 50 times | 48% |
| Oil temperature | TVD | SA | 54 times | 52% |
| Oil temperature | W | ES | 44 times | 42% |
| Oil temperature | TVD | ES | 63 times | 61% |
| Weather | W | SA | 10 times | 41% |
| Weather | TVD | SA | 13 times | 54% |
| Weather | W | ES | 9 times | 37% |
| Weather | TVD | ES | 9 times | 37% |
| NIFTY-50 | W | SA | 4 times | 33% |
| NIFTY-50 | TVD | SA | 6 times | 50% |
| NIFTY-50 | W | ES | 4 times | 33% |
| NIFTY-50 | TVD | ES | 5 times | 41% |

**Results**: Table 9 presents the reduced number of retraining in the model reuse approach compared to the periodically update approach. As seen in Table 9, on average, the model reuse approach reduces the number of retrainings by 29, 9, 52, 10, and 4 times for the NSW Electricity, Zurich Electricity, Oil Temperature, Weather, and NIFTY-50 datasets, respectively. Besides, Table 9 shows the percentage of reduced retraining count in the model reuse approach compared to the periodically update approach. As shown in Table 9, the model reuse approach reduces the number of retrainings by 46%, 36%, 50%, 42%, and 39% for the NSW Electricity, Zurich Electricity, Oil Temperature, Weather, and NIFTY-50 datasets, respectively, compared to the periodic model update approach. These results show that we can save a considerable amount of time and cost for the model update approach using our previously trained models.

Table 10 presents the computation time and cost of model reuse and the baseline approaches. We rescaled all the computation time and cost results over one year of using each method. The two highest computation costs are highlighted in the Table 10. As seen in the Table 10, the periodical retraining approach has the highest computaion cost. The informed model update has the second-highest computation cost when using RF and XGB learning algorithms.

Table 10: Comparison of the computaion time and costs of different model update approaches. The first two highest costs in each row are highlighted in the table. Stationary, periodical, reuse, and informed columns are different studies model update approaches. The time and cost units are seconds and US dollars, respectively.

| Dataset | Algorithm | Stationary | | Periodical | | Reuse | | Informed | |
|---|---|---|---|---|---|---|---|---|---|
| | | Time | Cost | Time | Cost | Time | Cost | Time | Cost |
| NSW Electricity | LSTM | 3.7 | $118 \times 10^{-6}$ | 8,62.3 | $\mathbf{27{,}507 \times 10^{-6}}$ | 1,06.5 | $\mathbf{3{,}397 \times 10^{-6}}$ | 65.7 | $2{,}095 \times 10^{-6}$ |
| Zurich Electricity | LSTM | 10.6 | $338 \times 10^{-6}$ | 13,41.9 | $\mathbf{42{,}806 \times 10^{-6}}$ | 1,43.6 | $4{,}580 \times 10^{-6}$ | 1,96.8 | $\mathbf{6{,}277 \times 10^{-6}}$ |
| Oil temperature | LSTM | 2.6 | $82 \times 10^{-6}$ | 5,81.4 | $\mathbf{18{,}546 \times 10^{-6}}$ | 1,91.4 | $\mathbf{6{,}105 \times 10^{-6}}$ | 1,12.2 | $3{,}579 \times 10^{-6}$ |
| Weather | LSTM | 19.9 | $634 \times 10^{-6}$ | 21,02.7 | $\mathbf{67{,}076 \times 10^{-6}}$ | 1,74.2 | $5{,}556 \times 10^{-6}$ | 2,26.1 | $\mathbf{7{,}212 \times 10^{-6}}$ |
| NIFTY-50 | LSTM | 0.4 | $12 \times 10^{-6}$ | 4,64.3 | $\mathbf{14{,}811 \times 10^{-6}}$ | 1.8 | $\mathbf{57 \times 10^{-6}}$ | 1.5 | $47 \times 10^{-6}$ |
| NSW Electricity | GRU | 4.2 | $133 \times 10^{-6}$ | 7,92.3 | $\mathbf{25{,}274 \times 10^{-6}}$ | 1,06 | $\mathbf{3{,}381 \times 10^{-6}}$ | 1,04.9 | $3{,}346 \times 10^{-6}$ |
| Zurich Electricity | GRU | 11.9 | $379 \times 10^{-6}$ | 1,037.4 | $\mathbf{33{,}093 \times 10^{-6}}$ | 1,44.9 | $4{,}622 \times 10^{-6}$ | 1,68.9 | $\mathbf{5{,}387 \times 10^{-6}}$ |
| Oil temperature | GRU | 2.2 | $70 \times 10^{-6}$ | 99.8 | $\mathbf{3{,}183 \times 10^{-6}}$ | 23 | $\mathbf{733 \times 10^{-6}}$ | 12.4 | $395 \times 10^{-6}$ |
| Weather | GRU | 21.2 | $676 \times 10^{-6}$ | 17,22.3 | $\mathbf{54{,}941 \times 10^{-6}}$ | 1,76.8 | $5{,}639 \times 10^{-6}$ | 2,39.4 | $\mathbf{7{,}636 \times 10^{-6}}$ |
| NIFTY-50 | GRU | 0.2 | $6.38 \times 10^{-6}$ | 3,45.7 | $\mathbf{11{,}027 \times 10^{-6}}$ | 1.5 | $\mathbf{47 \times 10^{-6}}$ | 1.2 | $38 \times 10^{-6}$ |
| NSW Electricity | RF | 0.5 | $15 \times 10^{-6}$ | 39.7 | $\mathbf{1{,}266 \times 10^{-6}}$ | 3.7 | $118 \times 10^{-6}$ | 9.5 | $\mathbf{303 \times 10^{-6}}$ |
| Zurich Electricity | RF | 3.9 | $124 \times 10^{-6}$ | 1,34.3 | $\mathbf{4{,}284 \times 10^{-6}}$ | 11.9 | $379 \times 10^{-6}$ | 23.9 | $\mathbf{762 \times 10^{-6}}$ |
| Oil temperature | RF | $\sim 0$ | $\sim 0$ | 28.3 | $\mathbf{902 \times 10^{-6}}$ | 3.9 | $124 \times 10^{-6}$ | 9.2 | $\mathbf{293 \times 10^{-6}}$ |
| Weather | RF | 1.3 | $41 \times 10^{-6}$ | 2,67.3 | $\mathbf{8{,}526 \times 10^{-6}}$ | 26.6 | $848 \times 10^{-6}$ | 47.8 | $\mathbf{1{,}524 \times 10^{-6}}$ |
| NIFTY-50 | RF | $\sim 0$ | $\sim 0$ | 9.7 | $\mathbf{309 \times 10^{-6}}$ | 0.1 | $3 \times 10^{-6}$ | 0.2 | $\mathbf{6 \times 10^{-6}}$ |
| NSW Electricity | XGB | $\sim 0$ | $\sim 0$ | 6.8 | $\mathbf{216 \times 10^{-6}}$ | 1 | $31 \times 10^{-6}$ | 9.5 | $\mathbf{303 \times 10^{-6}}$ |
| Zurich Electricity | XGB | $\sim 0$ | $\sim 0$ | 11.9 | $\mathbf{379 \times 10^{-6}}$ | 2.6 | $82 \times 10^{-6}$ | 14.6 | $\mathbf{465 \times 10^{-6}}$ |
| Oil temperature | XGB | $\sim 0$ | $\sim 0$ | 9.2 | $\mathbf{293 \times 10^{-6}}$ | 3.9 | $124 \times 10^{-6}$ | 9.9 | $\mathbf{315 \times 10^{-6}}$ |
| Weather | XGB | $\sim 0$ | $\sim 0$ | 11.9 | $\mathbf{379 \times 10^{-6}}$ | 1.3 | $41 \times 10^{-6}$ | 26.6 | $\mathbf{848 \times 10^{-6}}$ |
| NIFTY-50 | XGB | $\sim 0$ | $\sim 0$ | 0.3 | $\mathbf{9 \times 10^{-6}}$ | 0.1 | $3 \times 10^{-6}$ | 0.2 | $\mathbf{6 \times 10^{-6}}$ |

Table 11: Comparison of the average of the computation time and costs of different model update approaches for each dataset. The first two highest costs in each row are highlighted in the table. Stationary, periodical, reuse, and informed columns are different studied model update approaches. The time and cost units are seconds and US dollars, respectively.

| Dataset | Stationary | | Periodical | | Reuse | | Informed | |
|---|---|---|---|---|---|---|---|---|
| | Time | Cost | Time | Cost | Time | Cost | Time | Cost |
| NSW Electricity | 2.1 | $66 \times 10^{-6}$ | 4,25.2 | $\mathbf{13{,}565 \times 10^{-6}}$ | 54.3 | $\mathbf{1{,}731 \times 10^{-6}}$ | 47.4 | $1{,}511 \times 10^{-6}$ |
| Zurich Electricity | 6.6 | $210 \times 10^{-6}$ | 6,31.3 | $\mathbf{20{,}140 \times 10^{-6}}$ | 75.7 | $2{,}431 \times 10^{-6}$ | 1,09.0 | $\mathbf{3{,}222 \times 10^{-6}}$ |
| Oil temperature | 1.2 | $38 \times 10^{-6}$ | 1,79.6 | $\mathbf{5{,}731 \times 10^{-6}}$ | 55.5 | $\mathbf{1{,}771 \times 10^{-6}}$ | 41.3 | $1{,}145 \times 10^{-6}$ |
| Weather | 10.6 | $337 \times 10^{-6}$ | 10,26 | $\mathbf{32{,}730 \times 10^{-6}}$ | 94.7 | $3{,}021 \times 10^{-6}$ | 1,34.9 | $\mathbf{4{,}305 \times 10^{-6}}$ |
| NIFTY-50 | 0.15 | $4 \times 10^{-6}$ | 2,05 | $\mathbf{6{,}539 \times 10^{-6}}$ | 0.8 | $\mathbf{27 \times 10^{-6}}$ | 0.7 | $24 \times 10^{-6}$ |
| **Average** | 4.1 | $131 \times 10^{-6}$ | 4,93.4 | $\mathbf{15{,}741 \times 10^{-6}}$ | 56.2 | $1{,}796 \times 10^{-6}$ | 66.6 | $\mathbf{2{,}041 \times 10^{-6}}$ |

It also has the second-highest computation cost for the Zurich Electricity and the Weather datasets when using LSTM and GRU learning algorithms. These results show that the model reuse approach costs less than the periodical retraining approach using different configurations for all the datasets. It also costs less than the informed retraining for all the datasets when using the RF and XGB learning algorithms. It costs less than informed retraining for the Zurich Electricity and the Weather datasets when using LSTM and GRU learning algorithms.

Table 11 shows the average of the computaion time and costs of different model update approaches. Based on the results of Table 11, model reuse approach saves \$11834, \$17709, \$3960, \$29709, and \$6515 (all costs should be multiplied by $10^{-6}$) per year compared to the periodical retraining approach. Besides, based on the results of Table 11, the model reuse approach saves \$791, \$1284 per year for updating the model compared to the informed retraining approach for the Zurich Electricity and the Weather datasets (all costs should be multiplied by $10^{-6}$). Another interesting finding is that the informed model retraining approach costs more than the model reuse approach on average based on the results of Table 11. Furthermore, **the periodical and informed model retraining approaches take 8.7 and 1.13 times more computation time and cost than the model reuse approach, respectively.**

In some companies, training a model can take a long time (e.g., a few days or weeks) due to the complexity of the algorithm and the large volume of data. The longer the training takes, the more expensive it becomes. For example, imagine a scenario where training a model on ml.c5.24xlarge AWS SageMaker instance takes a week. In this case, the cost of periodically retraining the model will be \$822 each time, which is expensive. Therefore, we believe that our model reuse approach is beneficial for companies to save time and financial resources by reducing the number of unnecessary training. In addition, periodic and informed retraining come with extra costs that are not easy to measure. For example, training, testing, deploying new models, and detecting drift take up employees' time, which can become expensive over time.

It is worth mentioning that the model reuse approach will perform as well as a periodically updated model in the worst-case scenario. The worst-case scenario is where there is no similar distribution in the past time series dataset. In that case, the model reuse approach can not find a similar distribution in the past data, so it will retrain a new model on the most recent data. However, even in this scenario, the computation cost will not exceed the periodical retraining because forecasting and similarity measurement computation cost less than finding the proper segment length for the periodically updating model. Therefore, the model reuse approach is more suitable for the datasets with seasonality and recurrent patterns in the distribution.

It is essential to point out that storage of the model is costly. If we store 1GB of data on an AWS S3 bucket, it will cost \$0.023 per month (Amazon Web Services, 2025). The size of our biggest model is an RF model for the Weather dataset that takes 12MB of space on average. If we follow the model reuse approach, in the worst-case scenario where there is no similarity between each segment and its preceding segments, we should store 24 RF models per year, each of which takes 12MB on average. Thus, in total, the maximum required storage in this scenario would be 24 $\times$ 12MB, which is equal to a maximum of 288 MB at the end of one year. 228MB costs \$0.077 per year. Our approach needs the storage of historical models while the baseline approaches do not store the history of models.

**Summary of RQ2 results**

- Periodical retraining is the most costly model maintenance approach.
- The model reuse approach reduces the cost and time of model maintenance compared to periodic and informed retraining, while maintaining the performance comparable to both approaches.
- On average, the periodical and informed model retraining approaches take 8.7 and 1.13 times more time and cost than the model reuse approach, respectively.

## 7 Discussion and implication

This section further discusses the implications of our novel MLOps pipeline, model reuse approach, and SimReuse tool for companies, researchers, cloud service providers, and MLOps engineers.

**Companies and organizations**: Our RQ1 and RQ2 results show that the proposed pipeline and the SimReuse tool will help the companies reduce the number of unnecessary retraining, which leads to saved time, cost, and effort. Therefore, organizations, startups and companies can reduce the budget and time spent on retraining their models and invest time and money in more beneficial areas.

Lyu et al. (2021a) mentioned that periodical model retraining is the most frequently used model maintenance and concept drift adaptation method. However, the retraining model takes time and is costly. Some companies with large models and massive datasets may be unable to afford it. Besides, the stationary model's performance may degrade over time. Thus, our model reuse preserves the model performance while keeping the maintenance time and cost low for companies with low budgets and close deadlines.

**Researchers**: Preliminary study results show recurrent distribution patterns in time series datasets. Therefore, researchers can benefit from the approach and results of this study to find similar distribution patterns or to propose and develop tools that find similar data distributions in time series datasets and other types of datasets.

Furthermore, as we do not know the distribution of the actual upcoming data in the environment, it is critical to anticipate the distribution of the upcoming data in order to choose the right model to make predictions. Thus, researchers can benefit from our forecasting methods to find the distribution of their upcoming data. Moreover, they can propose new approaches to predict the distribution of the upcoming data. Finally, researchers and developers can provide visualization tools to monitor and compare the distributions of different time windows.

Besides, researchers can explore different forecasting and similarity measurement techniques to find which method works better and costs less for each data type.

**Cloud service providers**: This paper proposes an improved version of the MLOps pipeline. To facilitate the use of this pipeline, cloud service providers (e.g. AWS, Azure) can implement APIs to automate the recurrent distribution and seasonality detection in the datasets. Secondly, they can provide our improved MLOps pipeline in their systems for their users so that developers can easily use it in their projects on the cloud.

Our approach mentioned the importance of storing the trained models and reusing them for similar data distributions in the future. Therefore, cloud service providers can benefit from this work's findings by facilitating model storage and reuse for their users.

**MLOps engineers**: We proposed an improved MLOps pipeline in this study. Therefore, MLOps engineers can benefit from our improved MLOps pipeline to modify their model maintenance approach. They can integrate our MLOp pipeline to reduce their maintenance costs and time.

## 8 Threats to validity

This section explains the internal, external, construct, and conclusion validity of this study, along with the corresponding mitigation strategies.

### 8.1 Internal validity threats

One threat to the validity of this research is the effect of the methods used in this study on the results. We utilized two or more methods in each step to eliminate this threat. For example, to reduce the effect of the similarity measurement technique on the results, we used two different similarity metrics: Wasserstein distance and TVD. Another example is that we used two different learning algorithms to eliminate the effect of learning algorithms on our results. Therefore, the specific methods will not affect the generalizability of the results. Moreover, to find the proper segment length, we also used two methods to reduce each method's bias and effect. Finally, this study does not consider the storage cost because developers can save their models and data in their local storage.

Another threat to the validity of this research is finding the optimal segment length, which is key to the periodically updated model approach. Although we can test all the possible segment lengths for periodical retraining, it is costly and takes too much time. Therefore, it is not feasible to do that. To reduce the effort of finding the optimal segment length for 1) periodic training and 2) preliminary study and eliminate the risk of not considering all the possible values, we tested different values in the range of small segment lengths to large segment lengths.

### 8.2 External validity threats

The number of datasets used to evaluate the approach affects the generalizability of the results. To minimize this effect, we selected five widely used research datasets. We assume that our findings from these four time series datasets can be generalized to other time series datasets. However, our tool should be tested on time series datasets in other domains.

Besides, SimReuse is designed to reuse previously trained models when their training data distribution closely resembles that of the forecast data. However, this assumption may not apply to some scenarios. For example, in a holiday context, indeed, SimReuse can still rely on existing models if the conditions during the holiday match the distribution represented in the training data. However, some holidays may impact the distribution of some time series datasets dramatically. For instance, Black Friday will impact the web traffic of the Amazon website. Thus, SimReuse may not be effective enough on holidays like Black Friday and Boxing Day to forecast the web traffic.

### 8.3 Construct validity threats

One of the construct threats to validity is related to the design of the preliminary study. To find the proper segment length to split the different data distributions, we do a human analysis. Since human analysis is not free of mistakes, we conducted another analysis (model retraining) to find the proper segment length that minimizes the MSE. This way, we could select the final proper segment length confidently.

## 9 Conclusion and future work

This study explores the seasonality and recurrent drift in the time series datasets and proposes an improved MLOps pipeline and a new ML model maintenance approach that achieves comparable ML model performance to the baselines, while keeping the maintenance time and cost low. We propose a model reuse approach to maintain ML models in a production environment.

After conducting a preliminary study, we realized that there are seasonal and recurrent distribution patterns in the studied time series datasets. We benefit from the recurrent distribution patterns to avoid unnecessary periodical model retraining. Instead, we utilize our previously trained models for inference on the future data segments with distributions similar to our existing models' training sets. Our evaluation results show that our model reuse approach performs comparably to the periodical and informed model retraining approaches, and it decreases the training time and cost to 1/8th and 1/1.13th, respectively. Therefore, companies, MLOps engineers and developers can benefit from our improved MLOps pipeline and our model reuse approach to save time, cost and resources while maintaining their ML models.

Future research directions include extending our study beyond time series datasets to explore whether the model reuse approach can also be applied effectively to other data types, such as streaming data.

## 10 Declaration

**Ethical approval**: Not applicable. This study does not involve any human participants or animals. Therefore, we do not require ethical approval.

**Informed consent**: Not applicable. This study does not involve a survey or human participants. Therefore, we do not require informed consent.

**Author contributions**: Forough Majidi 1) proposed the idea and approach, and 2) contributed to writing the draft and resolving the writing issues. Foutse Khomh, Heng Li, and Amin Nikanjam contributed to 1) improving the idea and approach and 2) improving methods used in the study. They also contributed to improving the manuscript by changing the structure of the paper, detecting the writing issues, and proofreading it.

**Data availability**: All data used for analysis, as well as the scripts and the analysis results, are provided in (Majidi, 2024).

**Conflict of interest**: The authors declared that they have no conflict of interest.

**Clinical Trial Number**: Not applicable

**Appendix**

Tables 12, 13, 14, and 15 show the MSE and MAE performance results of the XGB and the GRU learning algorithms.

Table 12: Comparison of MSE of different model update approaches (learning algorithm: XGB)

| Dataset | Similarity metric | Forecast approach | Stationary model MSE | periodically updated model MSE | model reuse MSE | Informed retraining MSE | Random model MSE |
|---|---|---|---|---|---|---|---|
| NSW Electricity | W | SA | $16\times10^{-4}$ | $98\times10^{-5}$ | $85\times10^{-5}$ | $13\times10^{-4}$ | $34\times10^{-3}$ |
| NSW Electricity | TVD | SA | $16\times10^{-4}$ | $98\times10^{-5}$ | $49\times10^{-5}$ | $17\times10^{-4}$ | $34\times10^{-3}$ |
| NSW Electricity | W | ES | $16\times10^{-4}$ | $98\times10^{-5}$ | $35\times10^{-5}$ | $94\times10^{-5}$ | $34\times10^{-3}$ |
| NSW Electricity | TVD | ES | $16\times10^{-4}$ | $98\times10^{-5}$ | $95\times10^{-5}$ | $82\times10^{-5}$ | $34\times10^{-3}$ |
| Zurich Electricity | W | SA | $11\times10^{-3}$ | $67\times10^{-4}$ | $57\times10^{-4}$ | $99\times10^{-4}$ | $60\times10^{-3}$ |
| Zurich Electricity | TVD | SA | $11\times10^{-3}$ | $67\times10^{-4}$ | $76\times10^{-4}$ | $81\times10^{-4}$ | $60\times10^{-3}$ |
| Zurich Electricity | W | ES | $11\times10^{-3}$ | $67\times10^{-4}$ | $78\times10^{-4}$ | $71\times10^{-4}$ | $60\times10^{-3}$ |
| Zurich Electricity | TVD | ES | $11\times10^{-3}$ | $67\times10^{-4}$ | $69\times10^{-4}$ | $63\times10^{-4}$ | $60\times10^{-3}$ |
| Oil temperature | W | SA | $13\times10^{-2}$ | $59\times10^{-4}$ | $25\times10^{-4}$ | $83\times10^{-4}$ | $10\times10^{-3}$ |
| Oil temperature | TVD | SA | $13\times10^{-2}$ | $59\times10^{-4}$ | $35\times10^{-3}$ | $81\times10^{-4}$ | $10\times10^{-3}$ |
| Oil temperature | W | ES | $13\times10^{-2}$ | $59\times10^{-4}$ | $24\times10^{-4}$ | $10\times10^{-3}$ | $10\times10^{-3}$ |
| Oil temperature | TVD | ES | $13\times10^{-2}$ | $59\times10^{-4}$ | $88\times10^{-4}$ | $10\times10^{-3}$ | $10\times10^{-3}$ |
| Weather | W | SA | $13\times10^{-3}$ | $24\times10^{-5}$ | $13\times10^{-5}$ | $75\times10^{-5}$ | $37\times10^{-3}$ |
| Weather | TVD | SA | $13\times10^{-3}$ | $24\times10^{-5}$ | $11\times10^{-4}$ | $43\times10^{-5}$ | $37\times10^{-3}$ |
| Weather | W | ES | $13\times10^{-3}$ | $24\times10^{-5}$ | $22\times10^{-4}$ | $51\times10^{-5}$ | $37\times10^{-3}$ |
| Weather | TVD | ES | $13\times10^{-3}$ | $24\times10^{-5}$ | $29\times10^{-4}$ | $35\times10^{-5}$ | $37\times10^{-3}$ |
| NIFTY-50 | W | SA | $31\times10^{-4}$ | $14\times10^{-4}$ | $20\times10^{-4}$ | $25\times10^{-4}$ | $30\times10^{-3}$ |
| NIFTY-50 | TVD | SA | $31\times10^{-4}$ | $14\times10^{-4}$ | $32\times10^{-4}$ | $60\times10^{-4}$ | $30\times10^{-3}$ |
| NIFTY-50 | W | ES | $31\times10^{-4}$ | $14\times10^{-4}$ | $35\times10^{-4}$ | $64\times10^{-4}$ | $30\times10^{-3}$ |
| NIFTY-50 | TVD | ES | $31\times10^{-4}$ | $14\times10^{-4}$ | $35\times10^{-4}$ | $43\times10^{-4}$ | $30\times10^{-3}$ |

Table 13: Comparison of MAE of different model update approaches (learning algorithm: XGB)

| Dataset | Similarity metric | Forecast approach | Stationary model MAE | periodically updated model MAE | model reuse MAE | Informed retraining MAE | Random model MAE |
|---|---|---|---|---|---|---|---|
| NSW Electricity | W | SA | $25 \times 10^{-3}$ | $88 \times 10^{-4}$ | $94 \times 10^{-4}$ | $11 \times 10^{-3}$ | $10 \times 10^{-2}$ |
| NSW Electricity | TVD | SA | $25 \times 10^{-3}$ | $88 \times 10^{-4}$ | $89 \times 10^{-4}$ | $12 \times 10^{-3}$ | $10 \times 10^{-2}$ |
| NSW Electricity | W | ES | $25 \times 10^{-3}$ | $88 \times 10^{-4}$ | $81 \times 10^{-4}$ | $93 \times 10^{-4}$ | $10 \times 10^{-2}$ |
| NSW Electricity | TVD | ES | $25 \times 10^{-3}$ | $88 \times 10^{-4}$ | $10 \times 10^{-3}$ | $88 \times 10^{-4}$ | $10 \times 10^{-2}$ |
| Zurich Electricity | W | SA | $80 \times 10^{-3}$ | $59 \times 10^{-3}$ | $47 \times 10^{-3}$ | $73 \times 10^{-3}$ | $19 \times 10^{-2}$ |
| Zurich Electricity | TVD | SA | $80 \times 10^{-3}$ | $59 \times 10^{-3}$ | $57 \times 10^{-3}$ | $65 \times 10^{-3}$ | $19 \times 10^{-2}$ |
| Zurich Electricity | W | ES | $80 \times 10^{-3}$ | $59 \times 10^{-3}$ | $59 \times 10^{-3}$ | $61 \times 10^{-3}$ | $19 \times 10^{-2}$ |
| Zurich Electricity | TVD | ES | $80 \times 10^{-3}$ | $59 \times 10^{-3}$ | $48 \times 10^{-3}$ | $57 \times 10^{-3}$ | $19 \times 10^{-2}$ |
| Oil temperature | W | SA | $32 \times 10^{-2}$ | $59 \times 10^{-3}$ | $24 \times 10^{-3}$ | $71 \times 10^{-3}$ | $81 \times 10^{-3}$ |
| Oil temperature | TVD | SA | $32 \times 10^{-2}$ | $59 \times 10^{-3}$ | $85 \times 10^{-3}$ | $69 \times 10^{-3}$ | $81 \times 10^{-3}$ |
| Oil temperature | W | ES | $32 \times 10^{-2}$ | $59 \times 10^{-3}$ | $21 \times 10^{-3}$ | $80 \times 10^{-3}$ | $81 \times 10^{-3}$ |
| Oil temperature | TVD | ES | $32 \times 10^{-2}$ | $59 \times 10^{-3}$ | $62 \times 10^{-3}$ | $80 \times 10^{-3}$ | $81 \times 10^{-3}$ |
| Weather | W | SA | $65 \times 10^{-3}$ | $45 \times 10^{-4}$ | $28 \times 10^{-4}$ | $87 \times 10^{-4}$ | $15 \times 10^{-2}$ |
| Weather | TVD | SA | $65 \times 10^{-3}$ | $45 \times 10^{-4}$ | $93 \times 10^{-4}$ | $59 \times 10^{-4}$ | $15 \times 10^{-2}$ |
| Weather | W | ES | $65 \times 10^{-3}$ | $45 \times 10^{-4}$ | $15 \times 10^{-3}$ | $66 \times 10^{-4}$ | $15 \times 10^{-2}$ |
| Weather | TVD | ES | $65 \times 10^{-3}$ | $45 \times 10^{-4}$ | $15 \times 10^{-3}$ | $56 \times 10^{-4}$ | $15 \times 10^{-2}$ |
| NIFTY-50 | W | SA | $37 \times 10^{-3}$ | $17 \times 10^{-3}$ | $23 \times 10^{-3}$ | $30 \times 10^{-3}$ | $10 \times 10^{-2}$ |
| NIFTY-50 | TVD | SA | $37 \times 10^{-3}$ | $17 \times 10^{-3}$ | $36 \times 10^{-3}$ | $57 \times 10^{-3}$ | $10 \times 10^{-2}$ |
| NIFTY-50 | W | ES | $37 \times 10^{-3}$ | $17 \times 10^{-3}$ | $32 \times 10^{-3}$ | $68 \times 10^{-3}$ | $10 \times 10^{-2}$ |
| NIFTY-50 | TVD | ES | $37 \times 10^{-3}$ | $17 \times 10^{-3}$ | $26 \times 10^{-3}$ | $50 \times 10^{-3}$ | $10 \times 10^{-2}$ |

Table 14: Comparison of MSE of different model update approaches (learning algorithm: GRU)

| Dataset | Similarity metric | Forecast approach | Stationary model MSE | periodically updated model MSE | model reuse MSE | Informed retraining MSE | Random model MSE |
|---|---|---|---|---|---|---|---|
| NSW Electricity | W | SA | $24 \times 10^{-4}$ | $14 \times 10^{-4}$ | $13 \times 10^{-4}$ | $14 \times 10^{-4}$ | $34 \times 10^{-3}$ |
| NSW Electricity | TVD | SA | $24 \times 10^{-4}$ | $14 \times 10^{-4}$ | $11 \times 10^{-4}$ | $16 \times 10^{-4}$ | $34 \times 10^{-3}$ |
| NSW Electricity | W | ES | $24 \times 10^{-4}$ | $13 \times 10^{-4}$ | $11 \times 10^{-4}$ | $16 \times 10^{-4}$ | $34 \times 10^{-3}$ |
| NSW Electricity | TVD | ES | $24 \times 10^{-4}$ | $14 \times 10^{-4}$ | $15 \times 10^{-4}$ | $13 \times 10^{-4}$ | $34 \times 10^{-3}$ |
| Zurich Electricity | W | SA | $13 \times 10^{-3}$ | $11 \times 10^{-3}$ | $13 \times 10^{-3}$ | $13 \times 10^{-3}$ | $59 \times 10^{-3}$ |
| Zurich Electricity | TVD | SA | $13 \times 10^{-3}$ | $11 \times 10^{-3}$ | $21 \times 10^{-3}$ | $11 \times 10^{-3}$ | $59 \times 10^{-3}$ |
| Zurich Electricity | W | ES | $13 \times 10^{-3}$ | $11 \times 10^{-3}$ | $13 \times 10^{-3}$ | $13 \times 10^{-3}$ | $59 \times 10^{-3}$ |
| Zurich Electricity | TVD | ES | $13 \times 10^{-3}$ | $11 \times 10^{-3}$ | $14 \times 10^{-3}$ | $10 \times 10^{-3}$ | $59 \times 10^{-3}$ |
| Oil temperature | W | SA | $70 \times 10^{-3}$ | $60 \times 10^{-4}$ | $52 \times 10^{-4}$ | $10 \times 10^{-3}$ | $10 \times 10^{-3}$ |
| Oil temperature | TVD | SA | $70 \times 10^{-3}$ | $60 \times 10^{-4}$ | $28 \times 10^{-3}$ | $92 \times 10^{-4}$ | $10 \times 10^{-3}$ |
| Oil temperature | W | ES | $70 \times 10^{-3}$ | $61 \times 10^{-4}$ | $52 \times 10^{-4}$ | $72 \times 10^{-4}$ | $10 \times 10^{-3}$ |
| Oil temperature | TVD | ES | $70 \times 10^{-3}$ | $61 \times 10^{-4}$ | $10 \times 10^{-3}$ | $78 \times 10^{-4}$ | $10 \times 10^{-3}$ |
| Weather | W | SA | $10 \times 10^{-4}$ | $16 \times 10^{-5}$ | $17 \times 10^{-5}$ | $17 \times 10^{-5}$ | $37 \times 10^{-3}$ |
| Weather | TVD | SA | $10 \times 10^{-4}$ | $17 \times 10^{-5}$ | $24 \times 10^{-5}$ | $21 \times 10^{-5}$ | $37 \times 10^{-3}$ |
| Weather | W | ES | $10 \times 10^{-4}$ | $15 \times 10^{-5}$ | $29 \times 10^{-5}$ | $14 \times 10^{-5}$ | $37 \times 10^{-3}$ |
| Weather | TVD | ES | $10 \times 10^{-4}$ | $13 \times 10^{-5}$ | $41 \times 10^{-5}$ | $13 \times 10^{-5}$ | $37 \times 10^{-3}$ |
| NIFTY-50 | W | SA | $26 \times 10^{-4}$ | $97 \times 10^{-5}$ | $12 \times 10^{-4}$ | $16 \times 10^{-4}$ | $30 \times 10^{-3}$ |
| NIFTY-50 | TVD | SA | $26 \times 10^{-4}$ | $19 \times 10^{-4}$ | $31 \times 10^{-4}$ | $32 \times 10^{-4}$ | $30 \times 10^{-3}$ |
| NIFTY-50 | W | ES | $26 \times 10^{-4}$ | $12 \times 10^{-4}$ | $26 \times 10^{-4}$ | $27 \times 10^{-4}$ | $30 \times 10^{-3}$ |
| NIFTY-50 | TVD | ES | $26 \times 10^{-4}$ | $94 \times 10^{-5}$ | $10 \times 10^{-4}$ | $15 \times 10^{-4}$ | $30 \times 10^{-3}$ |

Table 15: Comparison of MAE of different model update approaches (learning algorithm: GRU)

| Dataset | Similarity metric | Forecast approach | Stationary model MAE | periodically updated model MAE | model reuse MAE | Informed retraining MAE | Random model MAE |
|---|---|---|---|---|---|---|---|
| NSW Electricity | W | SA | $34 \times 10^{-3}$ | $16 \times 10^{-3}$ | $16 \times 10^{-3}$ | $17 \times 10^{-3}$ | $10 \times 10^{-2}$ |
| NSW Electricity | TVD | SA | $34 \times 10^{-3}$ | $16 \times 10^{-3}$ | $17 \times 10^{-3}$ | $18 \times 10^{-3}$ | $10 \times 10^{-2}$ |
| NSW Electricity | W | ES | $34 \times 10^{-3}$ | $16 \times 10^{-3}$ | $16 \times 10^{-3}$ | $17 \times 10^{-3}$ | $10 \times 10^{-2}$ |
| NSW Electricity | TVD | ES | $34 \times 10^{-3}$ | $17 \times 10^{-3}$ | $19 \times 10^{-3}$ | $17 \times 10^{-3}$ | $10 \times 10^{-2}$ |
| Zurich Electricity | W | SA | $90 \times 10^{-3}$ | $83 \times 10^{-3}$ | $88 \times 10^{-3}$ | $90 \times 10^{-3}$ | $19 \times 10^{-2}$ |
| Zurich Electricity | TVD | SA | $90 \times 10^{-3}$ | $84 \times 10^{-3}$ | $11 \times 10^{-2}$ | $81 \times 10^{-3}$ | $19 \times 10^{-2}$ |
| Zurich Electricity | W | ES | $90 \times 10^{-3}$ | $83 \times 10^{-3}$ | $88 \times 10^{-3}$ | $90 \times 10^{-3}$ | $19 \times 10^{-2}$ |
| Zurich Electricity | TVD | ES | $90 \times 10^{-3}$ | $84 \times 10^{-3}$ | $94 \times 10^{-3}$ | $81 \times 10^{-3}$ | $19 \times 10^{-2}$ |
| Oil temperature | W | SA | $22 \times 10^{-2}$ | $60 \times 10^{-3}$ | $54 \times 10^{-3}$ | $78 \times 10^{-3}$ | $80 \times 10^{-3}$ |
| Oil temperature | TVD | SA | $22 \times 10^{-2}$ | $60 \times 10^{-3}$ | $10 \times 10^{-2}$ | $74 \times 10^{-3}$ | $80 \times 10^{-3}$ |
| Oil temperature | W | ES | $22 \times 10^{-2}$ | $60 \times 10^{-3}$ | $54 \times 10^{-3}$ | $67 \times 10^{-3}$ | $80 \times 10^{-3}$ |
| Oil temperature | TVD | ES | $22 \times 10^{-2}$ | $60 \times 10^{-3}$ | $78 \times 10^{-3}$ | $69 \times 10^{-3}$ | $80 \times 10^{-3}$ |
| Weather | W | SA | $21 \times 10^{-3}$ | $94 \times 10^{-4}$ | $92 \times 10^{-4}$ | $95 \times 10^{-4}$ | $15 \times 10^{-2}$ |
| Weather | TVD | SA | $21 \times 10^{-3}$ | $99 \times 10^{-4}$ | $10 \times 10^{-3}$ | $10 \times 10^{-3}$ | $15 \times 10^{-2}$ |
| Weather | W | ES | $21 \times 10^{-3}$ | $91 \times 10^{-4}$ | $11 \times 10^{-3}$ | $89 \times 10^{-4}$ | $15 \times 10^{-2}$ |
| Weather | TVD | ES | $21 \times 10^{-3}$ | $88 \times 10^{-4}$ | $13 \times 10^{-3}$ | $86 \times 10^{-4}$ | $15 \times 10^{-2}$ |
| NIFTY-50 | W | SA | $46 \times 10^{-3}$ | $22 \times 10^{-3}$ | $22 \times 10^{-3}$ | $28 \times 10^{-3}$ | $10 \times 10^{-2}$ |
| NIFTY-50 | TVD | SA | $46 \times 10^{-3}$ | $26 \times 10^{-3}$ | $45 \times 10^{-3}$ | $45 \times 10^{-3}$ | $10 \times 10^{-2}$ |
| NIFTY-50 | W | ES | $46 \times 10^{-3}$ | $23 \times 10^{-3}$ | $37 \times 10^{-3}$ | $43 \times 10^{-3}$ | $10 \times 10^{-2}$ |
| NIFTY-50 | TVD | ES | $46 \times 10^{-3}$ | $22 \times 10^{-3}$ | $21 \times 10^{-3}$ | $30 \times 10^{-3}$ | $10 \times 10^{-2}$ |